\documentclass[sigconf]{acmart}

\AtBeginDocument{%
  }

\copyrightyear{2026}
\acmYear{2026}
\setcopyright{cc}
\setcctype{by}
\acmConference[CHI '26]{Proceedings of the 2026 CHI Conference on Human Factors in Computing Systems}{April 13--17, 2026}{Barcelona, Spain}
\acmBooktitle{Proceedings of the 2026 CHI Conference on Human Factors in Computing Systems (CHI '26), April 13--17, 2026, Barcelona, Spain}
\acmPrice{}
\acmDOI{10.1145/3772318.3790458}
\acmISBN{979-8-4007-2278-3/2026/04}

\usepackage{longtable}
\usepackage{geometry}
\usepackage{pdflscape}
\usepackage{tabularx}
\usepackage{booktabs}
\usepackage{makecell}
\usepackage{array}
\usepackage{ragged2e} 
\usepackage{float}
\usepackage{multirow,booktabs,makecell}

\usepackage{xcolor} % Package for colored text
\definecolor{Issue1}{HTML}{B924A5} % magenta
\definecolor{Issue2}{HTML}{4860cc} % blue
\definecolor{Issue3}{HTML}{178F91} % emerald blue 
\definecolor{Issue4}{HTML}{08703D} % green
\definecolor{Issue5}{HTML}{6B753B} % olive green
\definecolor{Issue6}{HTML}{B9391D} % brick orange
\definecolor{Issue7}{HTML}{D76000} % orange
\definecolor{Issue8}{HTML}{6F3DB8} % pink
\definecolor{Issue9}{HTML}{08703D} % emerald green
\definecolor{Issue10}{HTML}{34217A} % purple
\definecolor{MISC}{HTML}{34217A} %868686

\renewcommand{\arraystretch}{1.05}

\newif\ifshowcommentsandchanges
\showcommentsandchangesfalse

\newcommand{\seungju}[1]{}
\newcommand{\hyehyun}[1]{}
\newcommand{\chen}[1]{}
\newcommand{\kain}[1]{}
\newcommand{\juho}[1]{}

\newcommand{\pointone}[1]{#1}
\newcommand{\pointtwo}[1]{#1}
\newcommand{\pointthree}[1]{#1}
\newcommand{\pointfour}[1]{#1}

\newcommand{\misc}[1]{#1}

\newcommand{\colorone}[1]{#1}
\newcommand{\colortwo}[1]{#1}

\newcommand{\colorfour}[1]{#1}
\newcommand{\colorfive}[1]{#1}

\newcommand{\colormisc}[1]{#1}

\ifshowcommentsandchanges
    \renewcommand{\seungju}[1]{{\color{blue}\bf{SJ: #1}\normalfont}}
    \renewcommand{\hyehyun}[1]{{\color{red}\bf{HH: #1}\normalfont}}
    \renewcommand{\chen}[1]{{\color{purple}\bf{CH: #1}\normalfont}}
    \renewcommand{\kain}[1]{{\color{orange}\bf{KA: #1}\normalfont}}
    \renewcommand{\juho}[1]{{\color{teal}\bf{JH: #1}\normalfont}}
    
    \renewcommand{\pointone}[1]{\colorbox{Issue1}{\textcolor{white}{\#1}}~\textcolor{Issue1}{#1}}
    \renewcommand{\colorone}[1]{{\color{Issue1}#1}}
    
    \renewcommand{\pointtwo}[1]{\colorbox{Issue2}{\textcolor{white}{\#2}}~\textcolor{Issue2}{#1}}
    \renewcommand{\colortwo}[1]{{\color{Issue2}#1}}
    
    \renewcommand{\pointthree}[1]{\colorbox{Issue3}{\textcolor{white}{\#3}}~\textcolor{Issue3}{#1}}

    \renewcommand{\pointfour}[1]{\colorbox{Issue4}{\textcolor{white}{\#4}}~\textcolor{Issue4}{#1}}
    \renewcommand{\colorfour}[1]{{\color{Issue4}#1}}

    \renewcommand{\colorfive}[1]{{\color{Issue5}#1}}

    \renewcommand{\misc}[1]{\colorbox{MISC}{\textcolor{white}{\#minor}}~\textcolor{MISC}{#1}}
    \renewcommand{\colormisc}[1]{{\color{MISC}#1}}
\fi

\begin{document}

\title[Accessibility Barriers in Video-Based Learning for Individuals with Borderline Intellectual Functioning]{``I Can’t Keep Up'': Accessibility Barriers in Video-Based Learning for Individuals with Borderline Intellectual Functioning}

\author{Hyehyun Chu}
\orcid{0009-0006-0256-5277} 
\affiliation{%
  \institution{School of Computing, KAIST}
  \city{Daejeon}
  \country{Republic of Korea}}
\email{hyenchu@kaist.ac.kr}

\author{Seungju Kim}
\orcid{0000-0001-7948-5909} 
\affiliation{%
  \institution{Information \& Electronics Research Institute, KAIST}
  \city{Daejeon}
  \country{Republic of Korea}}
\email{sjkim64891@kaist.ac.kr}

\author{Chen Zhou}
\orcid{0000-0003-1209-2590} 
\affiliation{%
 \institution{Computer Science, National University of Singapore}
 \city{Singapore}
 \country{Singapore}}
\email{zhouchen.zc@u.nus.edu}

\author{Yu-Kai Hung}
\orcid{0009-0003-1889-1630} 
\affiliation{%
 \institution{National Taiwan University}
 \city{Taipei}
 \country{Taiwan}}
\email{b09902040@csie.ntu.edu.tw}

\author{Saelyne Yang}
\orcid{0000-0003-1776-4712} 
\affiliation{%
  \institution{School of Computing, KAIST}
  \city{Daejeon}
  \country{Republic of Korea}}
\email{saelyne@kaist.ac.kr}

\author{Hyun W. Ka}
\orcid{0000-0002-1239-2502} 
\affiliation{%
  \institution{Assistive AI Lab, KAIST}
  \city{Daejeon}
  \country{Republic of Korea}}
\email{hyun.ka@kaist.ac.kr}

\author{Juho Kim}
\orcid{0000-0001-6348-4127}
\affiliation{
  \institution{School of Computing, KAIST}
  \city{Daejeon}
  \country{Republic of Korea}
}
\email{juhokim@kaist.ac.kr}
\affiliation{
  \institution{SkillBench}
  \city{Santa Barbara, CA}
  \country{USA}
}
\email{juho@skillbench.com}

\renewcommand{\shortauthors}{Chu et al.}

\newcommand{\revise}[1]{%
  \textcolor{red}{\fcolorbox{red}{red}{\textcolor{white}{R}}~#1}%
}

\ccsdesc[500]{Human-centered computing~Empirical studies in HCI}
\ccsdesc[500]{Human-centered computing~Empirical studies in accessibility}
\keywords{Video-Based Learning, Borderline Intellectual Functioning, Video Accessibility, Cognitive Accessibility, Inclusive Design}



\begin{abstract}

% The maximum is 150 words. Current total word count = 155 (https://wordcounter.net/)

Video-based learning (VBL) has become a dominant method for learning practical skills, yet accessibility guidelines provide limited guidance for users with cognitive differences.
In particular, challenges that individuals with Borderline Intellectual Functioning (BIF) encounter in video-based learning remain largely underexplored, despite VBL's potential to support their learning through features like self-paced viewing and visual demonstration.
To address this gap, we conducted a series of studies with BIF individuals and caretakers to comprehensively understand their VBL challenges.
Our analysis revealed challenges stemming from misalignment between user cognitive characteristics and video elements (e.g., overwhelmed by pacing and density, difficulty inferring omitted content), and experiential factors intensifying challenges (e.g., low self-efficacy).
\misc{While participants employed coping strategies such as repetitive viewing to address these challenges, these strategies could not overcome fundamental gaps with video.}
We further discuss the design implications on both content and UI-level features for BIF and broader groups with cognitive diversities.

\end{abstract}

\maketitle

\section{Introduction}

\pointone{
Video-based learning (VBL) has become a crucial component of enabling self-directed learning, especially for acquiring practical, everyday skills \cite{arkenback2025youtube, ng2014origins, pires2022learning}.
Videos offer rich visual demonstrations that make content more comprehensible compared to text-based resources \cite{mayer2002multimedia}.
To make VBL more accessible, prior accessibility research has explored ways to improve video learning experiences for people with low vision~\cite{liu21visual, liu22cross}, hearing impairments~\cite{liu22cross}, and cognitive disabilities~\cite{jiang2025shifting, zhu2025focusview}.
Among cognitively diverse groups, individuals with conditions such as ADHD often face substantial barriers when engaging with informational videos, including difficulties with sustained attention or processing complex information.
While prior work has proposed strategies to support such learners~\cite{zhu2025focusview}, a large population with Borderline Intellectual Functioning (BIF) has received far less attention.
}

Borderline Intellectual Functioning (BIF) refers to individuals with IQ scores ranging from 70-85, representing one type of cognitive differences \cite{fernell2020borderline}.
This population is substantial, representing approximately 13.6\% of the general population \cite{hassiotis2019association}.
\colorone{
However, due to cognitive characteristics such as working memory limitations \cite{alloway2010working} and difficulties with abstract reasoning \cite{jankowska2012strategies}, individuals with BIF often struggle with daily living tasks (e.g., financial management, navigating public transport) and learning occupational skills (e.g., following multi-step protocols, organizing tasks) \cite{sauter2023aaidd, hassiotis2019association, orio2025understanding, greenspan2017borderline}.
}
In addition, the diverse functional characteristics and co-occurring conditions within this population create additional challenges for accurate identification and support provision \cite{peltopuro2014borderline}.
Despite their significant presence and heterogeneous needs, this group remains underrepresented in accessibility research \cite{peltopuro2014borderline, sarsenbayeva2023mapping}.
In educational contexts, these learners often fall between support systems, where they are excluded from special education supports reserved for those with more severe disabilities, while simultaneously struggling in classrooms designed for neurotypical learners \cite{wieland2016time, segu2024routines, nouwens2020differentiating}.

In this context, VBL presents both opportunities and challenges for BIF learners.
On one hand, its inherent flexibility, such as repeated viewing, self-paced learning, and demonstration in multimodalities, could help address gaps in traditional educational settings \cite{schwan2004cognitive, barman_usefulness_2023}.
Indeed, VBL has been successfully applied in specialized contexts to teach self-management, hygiene, and academic skills to individuals with cognitive disabilities \cite{apa2013dsm5, mechling_ortega-hurndon_2007_cbvi, mechling2003multi_media}.
On the other hand, core cognitive characteristics of BIF, particularly working memory limitations, make it difficult to process the dense, multi-modal information typical for standard video formats \cite{apa2013dsm5, jiang202121}.
Since most instructional videos are not designed with BIF users in mind, their learning needs often go unmet.

Consequently, while VBL holds strong promise, the practices and challenges of BIF learners in this context remain underexplored.
Furthermore, existing web content accessibility guidelines such as WCAG \cite{W3C_WCAG22} primarily address sensory impairments, leaving cognitive differences like BIF underserved. 
This gap highlights the need for research that examines how VBL can be better designed to support the diverse cognitive and information-processing needs of BIF learners \cite{Shah2024_IJWesT, Acosta2020_IEEEAccess, Wu2023_IDD_Visualizations}.  

To address these gaps, we investigate the following research questions:
\begin{itemize}
    \item \textbf{RQ1}: How do different video elements and BIF users' cognitive traits impact video-based learning for BIF users?
    \item \textbf{RQ2}: What are the experiential factors that impact video-based learning for BIF users?
    \item \textbf{RQ3}: What strategies do BIF users employ during video-based learning, and what are their effectiveness and limitations?
\end{itemize}

We first conducted semi-structured interviews with social workers and parents of BIF individuals to understand the broader context of VBL challenges.
These interviews confirmed that VBL is essential for BIF individuals for acquiring critical life skills like emergency response training.
However, interviewees revealed that individuals with BIF often struggle to articulate their comprehension difficulties due to their limited meta-cognitive awareness \cite{bonifacci2008speed}.
Informed by these insights, we conducted a multi-methods study combining semi-structured interviews and observational video-watching with BIF users to investigate their experiences with VBL.
We first identified their self-reported challenges through a semi-structured interview and extended the findings through an observational study where they watched an educational video about \textit{how to use an AED (Automated External Defibrillator)} and took comprehension quizzes.

Regarding RQ1, we found that participants face challenges across multiple cognitive domains:
\textbf{Limited spatial perception} that hindered translation of graphic into 3D actions, \textbf{verbal comprehension challenges} that made even common terminology inaccessible, \textbf{working memory constraints} leading to cognitive overload from rapid pacing and information density, and \textbf{inferential reasoning difficulties} when extracting implicit information from dialogue or connecting omitted scenes.
\misc{For RQ2, these challenges were amplified by experiential factors, such as \textbf{diminished self-efficacy} from continuous negative feedback cycles accumulated through societal interactions.}
% In terms of RQ3, BIF users universally employed repetitive viewing as their primary coping strategy.
% However, while repetitive viewing offered psychological comfort, it caused users to repeatedly encounter the} \pointfive{same comprehension barriers without resolution}.
In terms of RQ3, we found that BIF users employed limited coping strategies when encountering comprehension difficulties.
While they readily sought help for technical issues, they \textbf{remained silent about content comprehension challenges} and instead relied primarily on repetitive viewing.
However, while repetitive viewing offered psychological comfort, it caused users to \textbf{repeatedly encounter the same comprehension barriers} without resolution.

By understanding specific accessibility barriers through interviews and observed behaviors, we discuss design implications for BIF-inclusive video learning.
In summary, our contributions are: (1) insights into BIF users' challenges and current strategies in video-based learning through an interview and observational study, and (2) design implications for more inclusive video experiences, such as reducing cognitive load and fostering self-efficacy of BIF users.

\section{Related Work}

\subsection{Understanding Borderline Intellectual Functioning}
Borderline Intellectual Functioning (BIF) describes individuals with mild cognitive impairments that affect academic performance, social functioning, and independent living skills \cite{salvador2013borderline}.
While the fourth edition of the Diagnostic and Statistical Manual of Mental Disorders (DSM-IV) defined BIF by IQ scores of 70-85 \cite{bell1994dsm}, DSM-V eliminated this numerical threshold and now requires assessment of adaptive functioning difficulties \cite{apa2013dsm5}.
This shift acknowledges that individuals with BIF experience substantial daily difficulties despite having cognitive abilities above the intellectual disability threshold \cite{salvador2013borderline, sauter2023aaidd, lee2024epidemiology}.
However, the absence of clear diagnostic criteria has created identification challenges, leaving this population with limited access to appropriate support services \cite{wieland2016time, hassiotis2022borderline}.

Individuals with BIF often appear to function without difficulty, but this perception is misleading.
Research indicates that individuals with BIF actively attempt to conceal their need for assistance and mask their challenges \cite{wieland2015psychopathology}.
While earlier diagnostic standards (DSM-III and DSM-IV) suggested that these individuals do not exhibit \colorfive{difficulties} in adaptive behavior \cite{elsevier_dsm3r, bell1994dsm}, subsequent research has revealed a considerably different reality.
Previous studies indicate that individuals with BIF experience various daily living difficulties compared to the neurotypical population \cite{salvador2013borderline, lee2024epidemiology, BIFConsensusGroup2017Recommendations}.
The diverse functional characteristics and co-occurring conditions within this population create additional challenges for accurate identification and support provision \cite{peltopuro2014borderline}.

Research has identified consistent cognitive characteristics defining the BIF population.
These individuals demonstrate working memory \colorfive{constraints} in both verbal and visuo-spatial domains, which significantly affect their learning capacity and daily problem-solving abilities \cite{alloway2010working, henry2001does}.
Social abilities are particularly impaired, with difficulties in perspective-taking and social situation interpretation linked to inadequate social information processing systems \cite{van2011development}.
These cognitive limitations result in similar psychosocial and adaptive functioning \colorfive{challenges} to those with mild intellectual disability \cite{sauter2023aaidd, lee2024epidemiology}.

However, the borderline nature of this population creates unique systemic barriers that extend beyond cognitive limitations alone.
The absence of clear diagnostic criteria creates a legislative ``gray area'' where individuals with BIF fall between eligibility thresholds–too high-functioning for disability services yet unable to succeed with mainstream resources \cite{BIFConsensusGroup2017Recommendations, segu2024routines, nouwens2020differentiating}.
According to recent ethnographic research \cite{segu2024routines}, absence of appropriate support systems forces families of BIF people into the desperate position of overstating their children's challenges to access necessary disability services.
Their ethnographic findings reveal that this systemic exclusion creates reinforcing cycles of disadvantage across educational, employment, and social contexts, ultimately transforming manageable cognitive differences into significant barriers to independence and community participation.

\subsection{The Promise of Video-based Learning for the BIF population}
Given the working memory \colorfive{constraints} and the slower processing speeds of BIF, video-based learning offers particular promise to support this population.
According to the Cognitive Theory of Multimedia Learning \cite{mayer2002multimedia}, learners process visual and auditory information through separate working memory channels.
This dual-channel processing can particularly benefit individuals with BIF, who face cognitive bottlenecks when processing complex information \cite{hecker2002benefits, sweller1999instructional}.
\pointthree{
Prior accessibility research supports this by demonstrating that multimodal interfaces effectively reduce cognitive load for users with cognitive disabilities \cite{van2018comparing, koushik2022towards}, highlighting the potential benefits of video's multimodal presentation.
}

Thoughtfully designed videos minimize the cognitive effort required to understand procedures.
Video's sequential demonstrations reduce intrinsic cognitive load by making abstract concepts concrete \cite{sweller1999instructional}.
Specific features could potentially support BIF's cognitive profile: closed captions provide redundant information to aid comprehension \cite{jones2007symbols}, picture symbols reduce reliance on abstract reasoning \cite{detheridge2013literacy, slater2002pictorial}, and replay functionality accommodates slower processing speeds.
This redundant cross-modal presentation could be particularly valuable for BIF learners with information retention difficulties \cite{evmenova2011effects, hecker2002benefits}.
Additionally, appropriate pacing minimizes extrinsic load \cite{mayer2024past} while dual channels circumvent literacy barriers \cite{clark_paivio_dual_coding_1991, mayer_heiser_lonn_cognitive_constraints_2001}.

Educational research has established video-based instruction (VBI) as an effective approach for various populations with cognitive disabilities, such as individuals with intellectual disabilities \cite{wright2020review, nosrati2025combining}, autism spectrum disorders \cite{ALEXANDER20131346, allen2010use, van2010comparison, wright2020review}, and learning disabilities across daily living skills, vocational training, and academic domains \cite{ALEXANDER20131346, mechling2007computer, marzullo2011using, mechling2011computer}.
VBI provides clear demonstrations, allowing learners to learn through mimicking \cite{bandura1965influence} and simulate skills in various settings \cite{mechling2007computer}.
Established approaches like video modeling (demonstrating behaviors for imitation) and video prompting (breaking skills into sequential steps) have demonstrated success across related populations \cite{bandura1965influence, wright2020review}, suggesting similar potential for individuals with BIF. 

However, this effectiveness primarily comes from structured instructional settings with specialized content and support.
In contrast, independent learning from mainstream video content—such as informational videos on online video platforms like YouTube—presents different accessibility challenges.

\subsection{Why Current Video Accessibility Fails the BIF Population}
Despite VBL's promise for individuals with BIF, current accessibility frameworks and video design practices create significant barriers that prevent this population from realizing these benefits.
The fundamental challenge lies in how existing digital accessibility standards approach cognitive disabilities compared to sensory impairments.

Current accessibility guidelines, such as the Web Content Accessibility Guidelines (WCAG) \cite{W3C_WCAG22, Shah2024_IJWesT}, were primarily designed for sensory and physical disabilities and lack specific guidance for the diverse cognitive and information processing challenges characteristic of BIF \cite{Shah2024_IJWesT, Acosta2020_IEEEAccess, Wu2023_IDD_Visualizations}.
This creates a critical mismatch: while sensory accessibility focuses on translating information between modalities (e.g., visual to auditory through captions), cognitive accessibility requires careful management of information processing demands \cite{BernabeCaro2020_TaxonomyE2U}.

For individuals with BIF, whose working memory \colorfive{constraints} and slower processing speeds were established in previous research \cite{alloway2010working, henry2001does}, this distinction becomes particularly problematic.
While prior work has explored how multimodal translations can support neurodivergent groups like ADHD by allocating cognitive resources \cite{jiang2025shifting, zhu2025focusview}, traditional accessibility features can create complexity of help where accommodations designed for one group may inadvertently increase cognitive load for another.
For instance, captions that must be read concurrently with watching visual demonstrations can split attention between multiple information streams, potentially overwhelming the limited working memory capacity characteristic of BIF \cite{Anagha2023_VideoCaptionAccessibility, Sloan2006_MultimediaAccessibility}.
While some BIF individuals benefit greatly from textual reinforcement, others find dual-channel processing overwhelming, highlighting the need for flexible, user-controlled solutions.

This cognitive vulnerability is compounded by poor interface design.
Even when accessibility features are available, they are often buried in complex menus that require multi-step, cognitively demanding setup processes \cite{Nielsen2012_Usability101}.
The mental effort required to simply locate and activate accessibility features imposes additional extrinsic cognitive load, often defeating the feature's purpose before it can be utilized.

\pointthree{
While mainstream interfaces often fail to support these needs, prior work on accessible interfaces has demonstrated that thoughtfully designed interaction patterns can effectively scaffold task performance.
Studies on simplified user interfaces, structured prompting systems, and AR coaching for people with intellectual and cognitive disabilities demonstrate how clearly indicating ``what to do next'' can enable users to carry out complex multi-step routines with support \cite{esposito2024advancing, philips2024helping, koushik2022towards}.
Similarly, research on assistive technology for people with sensory impairments that support user-driven customization illustrates how users or their supporters can configure highly personalized support tools that fit an individual's specific routines and preferences \cite{herskovitz2024diy}.
A recent scoping review further highlights the breadth of such specialized tools developed for populations with diagnosed intellectual disabilities such as Down syndrome \cite{l2025accessible}.}

\pointone{However, the diverse cognitive nature and social context of BIF create a unique challenge that these current interventions overlook.
Most existing interventions assume substantial configuration effort from caregivers or the users themselves.
This configuration burden is particularly problematic for BIF users.
Unlike populations with diagnosed Intellectual and Developmental Disabilities (IDD) who can benefit from caregiver networks to manage customized assistive technology \cite{koushik2022towards}, BIF individuals typically manage devices independently.
In this independent context, they can struggle to navigate and understand the complex, abstract menus typical of accessibility settings.
Current interfaces lack the \textit{concrete scaffolding} which is necessary to support their decision-making, which can overwhelm their working memory limits and lead to abandonment of potentially helpful features.
Furthermore, individuals with BIF often strive to mask their difficulties to avoid stigma \cite{wieland2015psychopathology}, which can lead to a rejection of specialized technologies that identify them as ``disabled'' \cite{philips2024helping}.
Consequently, current approaches fall short for this ``in-between'' group: existing DIY tools are too cognitively demanding to be used independently, while specialized assistive technology solutions can be perceived as too stigmatizing.
}

The BIF population's unique position between neurotypical expectations and intellectual disability creates additional design challenges that existing frameworks inadequately address.
Current accessibility approaches lack the flexibility needed to accommodate the diverse functional characteristics within this population \cite{peltopuro2014borderline}, highlighting a critical gap in accessible video design that this study aims to bridge.

\section{Methodology}

To comprehensively understand VBL challenges for BIF users, we conducted a two-part study.
We first interviewed social workers and parents of BIF individuals (Section~\ref{method-SWParent}) to understand the broader context of VBL challenges and gather study design insights.
These interviews revealed that VBL is essential for acquiring critical life skills (e.g., fire extinguisher) and highlighted that BIF individuals' limited meta-cognitive awareness~\cite{bonifacci2008speed} often prevents them from articulating comprehension difficulties.
Informed by these insights, we conducted our main study with 12 BIF users (Section~\ref{method-BIF}) combining semi-structured interviews about VBL experiences with observational video-watching sessions where participants completed comprehension quizzes.
\pointfour{This multimethod approach~\cite{morse2003principles} allowed us to capture both conscious experiences through interviews and implicit comprehension barriers through observation.
During analysis, we prioritized BIF users' direct experiences as our primary data source.
Insights from caregivers and professionals served as a secondary layer for contextualization, ensuring the findings remained grouped in BIF users' lived experiences while accounting for broader systemic challenges.}

\subsection{Interviews with Social Workers and Parents with BIF Children} \label{method-SWParent}

We conducted interviews with social workers and parents of BIF individuals who regularly observe BIF individuals' video learning behaviors and could provide insights into their real-world VBL experiences.
We recruited social workers as representatives for the most diverse BIF user case observers and parents as the most frequent observers.

\subsubsection{Participants}

\begin{table*}[htbp]
\centering
\caption{Interview Participants: Social Workers. Professional profiles of two social workers who participated in interviews. Both participants have extensive experience supporting individuals with BIF.}
\label{tab:socialworkers}
\begin{tabularx}{\textwidth}{l l X}
\toprule
\textbf{Participant} & \textbf{Professional Background} & \textbf{BIF Clients Supported (approx.)} \\
\midrule
SW 1 & 17+ years; leader of BIF specialized programs; & $\sim$200+ across children, adolescents, and adults (up to age 39) \\
\midrule
SW 2 & 8+ years; Social Welfare major; & $\sim$50+ (slow learners, unregistered and multicultural youth) \\
\bottomrule
\end{tabularx}
\end{table*}

\begin{table*}[htbp]
\centering
\caption{Interview Participants: Parents. Demographic profiles and contextual information of children with BIF, as reported by four parents to inform the main study design.}
\label{tab:parents}
\begin{tabularx}{\textwidth}{l l l X}
\toprule
\textbf{Participant} & \textbf{Child Profile} & \textbf{Child's Diagnosis} & \textbf{Reported Characteristics and Video Behaviors} \\
\midrule
Parent 1 & \makecell[tl]{Young adult, \\ Gender unknown} & BIF & 
Struggles with abstract narratives in videos; challenges in daily independence; requires ongoing support for life management. \\
\midrule
Parent 2 & \makecell[tl]{Middle school, \\ Gender unknown} & BIF, ADHD & 
High proficiency in smartphone/YouTube use; learns origami through repetitive video playback; struggles with sustained attention except when highly interested. \\
\midrule
Parent 3 & \makecell[tl]{Middle school,\\ Male} & \makecell[tl]{BIF, Dyslexia, \\ ADHD, anxiety}& 
Delayed fine-motor skills; limited mathematical comprehension; uses motivational videos to enhance self-esteem; vulnerable to peer interactions. \\
\midrule
Parent 4 & \makecell[tl]{Grade 9, Male} & BIF & 
Retains only 20--30\% of learned information; performs at elementary level; cannot follow video lectures but learns autonomously through interest-driven content (e.g., basketball videos). \\
\bottomrule
\end{tabularx}
\end{table*}

We conducted interviews with two experienced social workers and four parents of individuals with BIF (see Table~\ref{tab:socialworkers} and Table \ref{tab:parents}).
Social workers were recruited through direct contact with welfare facilities and disability support organizations.
Participants had 8 and 17 years of experience supporting BIF populations and provided expert insights on learning patterns and video accessibility challenges observed in their professional practice.

Parent participants were recruited through snowball sampling \cite{naderifar2017snowball} facilitated by the social worker participants due to difficulty of recruiting.
These parents served as representatives of regional BIF caregiver support groups in South Korea's public communities.
They were expected to provide insights informed not only by their own children's experiences but also by their regular observations of learning patterns and challenges across multiple BIF families within support networks.

\subsubsection{Procedure}

The study session lasted approximately 60 minutes and was conducted online over Zoom.
Each session followed a structured protocol and explored: (1) learning characteristics of BIF individuals and observed specific challenges and barriers during VBL, (2) observed patterns of how BIF individuals or child interact with video content, (3) opinions on existing video interaction, (4) contexts where VBL becomes essential for BIF individuals, and (5) feature suggestion for VBL.
The detailed interview questionnaires can be found in Appendix~\ref{appendix-SWParentInterview}.
Social workers were compensated 250,000 KRW ($\approx$ 180 USD) and parents participants were compensated 50,000 KRW ($\approx$ 38 USD) for their participation in the study.

Specifically, these interviews revealed that BIF individuals are often unable to articulate when or why their comprehension breaks down during video watching, highlighting the need for observational methods and structured comprehension assessments to complement 
self-reported difficulties.
In addition, parents emphasized the critical importance of VBL for acquiring life-saving skills (e.g., AED use, fire extinguisher operation), which informed our selection of an AED instructional video as the study material. 
In the Results section (Section~\ref{result}), we report the findings together as a support for the result of thematic analysis of study with BIF users.

\subsection{Study with BIF Users} \label{method-BIF}

Based on insights from the interviews, we designed our study protocol with BIF users.
To investigate the accessibility challenges experienced by individuals with BIF when learning from informational videos, we conducted semi-structured interviews and video-watching sessions with 12 participants about their experiences with VBL and how-to videos.

\subsubsection{Participants}

\begin{table*} [htbp]
\centering
\caption{Participant demographics and functional characteristics. Demographic information includes participant ID, gender, age, intellectual disability diagnosis (MID = Mild Intellectual Disability), IQ score, and comorbid conditions. Functional domains assessed include: \textit{Overall} (general difficulty level across all domains), \textit{Attention} (ability to sustain focus during video-watching sessions), \textit{Memory} (recall of video content and past experiences), \textit{Dialogue} (fluency and ease in verbal interaction), \textit{Work} (difficulties experienced on employment), and \textit{School} (difficulties experienced in formal education). Difficulty levels are coded as: X = no difficulty, $\triangle$ = mild difficulty, and O = severe difficulty. Dashes (--) indicate absence of formal diagnosis.}
\label{tab:participant}
\begin{tabularx}{\textwidth}{X c c c c c cccccc}
\toprule
\textbf{ID} & \textbf{Gender} & \textbf{Age} & \textbf{Disability} & \textbf{IQ} & \textbf{Comorbidity} & \textbf{Overall} & \textbf{Attention} & \textbf{Memory} & \textbf{Dialogue} & \textbf{Work} & \textbf{School} \\
\midrule
P1  & Female            & 26 & --            & --          & --    
    & \(\triangle\) & \(\triangle\) & X & X & \(\triangle\) & \(\triangle\) \\
\midrule
P2  & Male              & 29 & MID     & 71          & --    
    & O & \(\triangle\) & X & \(\triangle\) & O & O \\
\midrule
P3  & Male              & 23 & --            & 82          & --    
    & \(\triangle\) & \(\triangle\) & O & X & \(\triangle\) & \(\triangle\) \\
\midrule
P4  & \makecell{Prefer not\\to say}              & 28 & --            & 64          & Anxiety 
    & O & \(\triangle\) & O & \(\triangle\) & O & O \\
\midrule
P5  & Female & 26 & --            & 75          & Depression 
    & O & O & \(\triangle\) & O & O & O \\
\midrule
P6  & Male & 27 & --      & --          & --    
    & O & O & O & O & O & O \\
\midrule
P7  & Male & 27 & --     & 82      & --    
    & \(\triangle\) & \(\triangle\) & \(\triangle\) & X & \(\triangle\) & \(\triangle\) \\
\midrule
P8  & Male  & 27 & MID            & 69-71          & Depression 
    & \(\triangle\) & \(\triangle\) & X & X & \(\triangle\) & O \\
\midrule
P9  & Male & 30 & --            & -           & Anxiety 
    & \(\triangle\) & O & O & \(\triangle\) & \(\triangle\) & \(\triangle\) \\
\midrule
P10 & Female& 25 & --            & -           & --    
    & \(\triangle\) & \(\triangle\) & X & X & \(\triangle\) & O \\
\midrule
P11 & Male& 27 & --            & -           & --    
    & O & O & O & O & \(\triangle\) & O \\
\midrule
P12 & Male& 33 & --            & 75          & \makecell{ADHD, Dyslexia, \\ Depression, Anxiety, \\ Panic Disorder} 
    & \(\triangle\) & O & \(\triangle\) & X & \(\triangle\) & \(\triangle\) \\
\bottomrule
\end{tabularx}
\end{table*}

We recruited 12 participants who had either been formally diagnosed with BIF, or people who received support services designed for BIF.
Participants were recruited through a snowballing method \cite{naderifar2017snowball} based on disability support services, community centers, and online forums focused on cognitive accessibility.
The inclusion criteria required participants to be adults (19+ years) and have experiences with VBL.
Additionally, participants needed to meet at least one of the following BIF identification criteria: an IQ score in the range of 64-84, a formal diagnosis of BIF from medical professionals, or documented experience receiving support services for BIF from welfare organizations. 
This functionality-based approach was chosen to emphasize the environmental and experiential aspects of cognitive accessibility challenges rather than relying solely on medical classification.
During recruitment, we conducted screening through a survey that collected participants' self-reported VBL experience.

Participant demographics and diagnostic information are summarized in Table \ref{tab:participant}.
We recruited 12 participants (8 male, 3 female, 1 undisclosed) aged 23-33 years with IQ scores ranging from 64-82.
Participants were compensated 50,000 KRW ($\approx$ 38 USD) per hour for their participation in the study.
Our study procedure was approved by the Institutional Review Board (IRB) of our institution.

\subsubsection{Procedure}
The study was conducted online and lasted approximately 60 to 90 minutes. Each session followed a structured protocol that consisted of two main parts. The first part was a 30--minute semi-structured interview about general VBL experiences and challenges. This was followed by a 30--60 minute video-watching session.

\paragraph{Semi-Structured Interview about VBL experiences}
The first part of the session explored participants' experiences with VBL.
We began with general background questions about learning difficulties in formal and informal contexts, then examined video consumption patterns with questions such as ``How often do you watch videos in a week?''
\pointfour{We then focused on informational video experiences, and explored how they find and select videos (e.g., ``How do you usually search for videos that suit your needs?'') and their memorable experiences—both positive and challenging (e.g., Could you share a specific example of the informational video that was difficult to follow?'').}
We investigated specific barriers based on the experiences that participants shared.
Throughout the interview, participants described coping strategies they had developed and provided insights into elements that either facilitated or hindered their learning process.
The detailed interview questionnaires can be found in Appendix~\ref{appendix-BIFInterview}.

\paragraph{Video-Watching Session with Quiz}
To address the limitations of self-reported data and systematically examine how specific video elements affect comprehension for individuals with BIF \cite{bonifacci2008speed}, we conducted a video-watching session where participants watched an instructional video and solved quizzes designed by the researchers, while researchers observed their comprehension process in real-time.
This approach was designed to capture real-time struggles that participants might not be able to articulate or remember during interviews.

Based on insights from interviews with parents, who emphasized the importance of VBL for acquiring essential life-saving skills, we selected an Automated External Defibrillator (AED) instructional video. 
We selected a recently produced AED video from the Korea Disease Control and Prevention Agency\footnote{\href{https://www.youtube.com/watch?v=b4XAPbxcEZk}{Korea Disease Control and Prevention Agency. 2023. Life-saving CPR 6. How to Use an Automated External Defibrillator (AED). YouTube.}}, which was designed for all citizens and represents typical government-produced instructional content that individuals with cognitive limitations can encounter.
This video was chosen through researcher discussion for its optimal length (approximately 2 minutes) and clear presentation approach, making it representative of government-produced alternatives while being suitable for our study purposes.

The selected AED instructional video was chosen for its typical instructional video characteristics that could potentially challenge BIF users: (1) a structured step-by-step format with 4 procedural stages, (2) dual-channel information delivery combining verbal narration with visual demonstrations, (3) use of simple but technical language, and (4) a standard pacing of approximately 2 minutes.
% These elements represent common features found in institutional-produced instructional content that individuals with cognitive limitations regularly encounter.
\colorfour{These elements represent typical design patterns in institutional-produced instructional content that can create accessibility barriers for individuals with cognitive limitations \cite{jiang202121, Sloan2006_MultimediaAccessibility}.}

To capture participants' actual struggles when actively trying to understand the content, we developed comprehension quizzes based on key video elements that could potentially hinder understanding.
This quiz-based approach was designed based on the prior research and the suggestion of social workers, as people with BIF often experience difficulties with self-monitoring abilities \cite{bonifacci2008speed}.
\colorfour{Drawing on WCAG video accessibility guidelines \cite{W3C_WCAG22}, we first segmented the video according to subtitle chapters, then systematically identified video elements that could create comprehension barriers.
Questions were designed to probe how these video elements interact with multiple cognitive abilities (e.g., spatial perception, working memory, inferential reasoning), as video comprehension inherently requires integrated processing across domains.}
The detailed questions and design rationales for each question can be found in Appendix~\ref{appendix-quiz}.

During the session, participants answered comprehension questions after each video segment while researchers documented moments where participants visibly struggled---such as confused expressions, replaying sections, or explicit statements of difficulty.
\colorfour{To examine detailed understanding and comprehension strategies for questions requiring multiple abilities, follow-up questions were used.
Interviewers also read questions aloud and provided explanations to prevent reading difficulties from hindering responses.}
Following each assessment, participants explained the confidence level about their answer, which specific aspects they found challenging, and why those elements were difficult to understand.
After these two sessions, the study was wrapped up with future suggestions for improving video accessibility.

\subsection{Data Analysis}
We employed reflective thematic analysis to analyze the interview data following the approach developed by Braun and Clarke~\cite{braun2019reflecting}.
Four researchers participated in the iterative coding process.
Initially, two researchers who had conducted the interviews independently coded three BIF participant interview transcripts along with observational field notes from the video-watching sessions.
These researchers engaged in deep data immersion to develop preliminary codes and an initial codebook through iterative discussion sessions.
Subsequently, the remaining two researchers applied this initial codebook to analyze the remaining nine BIF participant interviews.
All four researchers then gathered to discuss and merge coding disagreements, refine code definitions, and resolve conflicts.
In total, it took five iterations to finalize the codebook.

Our analysis prioritized identifying challenges experienced by BIF participants as the primary source of findings.
We employed this approach to ensure that our findings remained centered on the experiences of BIF users while using caregiver and professional observations to validate and contextualize these experiences.
Once themes emerged from the BIF participant study data, we systematically reviewed the social worker and parent interview transcripts to identify corroborating evidence rather than analyzing them as separate data sources.
Using the finalized codebook, the team reviewed social worker and parent interview transcripts to identify supporting evidence for themes that emerged from the BIF participant study analysis.
Quotes from social workers and parents were included when they provided: (i) observational evidence supporting participant-reported challenges, (ii) contextual insights explaining real-world implications, or (iii) professional perspectives on the broader impact of identified accessibility issues.

\section{Results} \label{result}
\begin{figure*}
    \centering
    \includegraphics[width=1.0\linewidth]{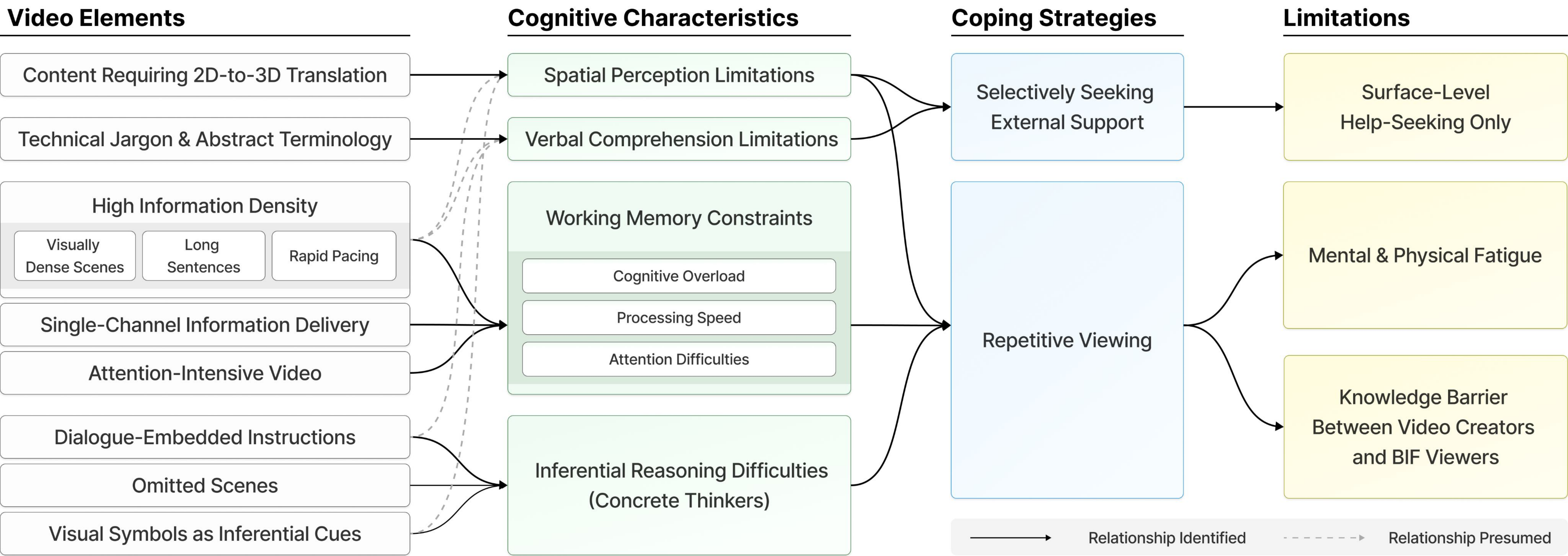}
    \caption{Overview of how video elements map to cognitive characteristics, coping strategies, and limitations for BIF users in video-based learning. 
    Left: video elements that heighten cognitive demand—content requiring 2D to 3D translation; technical jargon and abstract terminology; high information density (visually dense scenes, long sentences, rapid pacing); single-channel information delivery; attention-intensive video; dialogue-embedded instructions; omitted scenes; and visual symbols used as inferential cues. 
    Middle: cognitive characteristics—spatial perception and verbal comprehension limitations; working-memory constraints (cognitive overload, processing speed, attention difficulties); and inferential-reasoning difficulties among concrete thinkers. 
    Right: coping strategies (seeking external support, repetitive viewing) and resulting limitations (asynchronous help-seeking deadlock; mental and physical fatigue; knowledge barrier between video creators and BIF viewers). 
    Solid connectors denote relationships identified in our data; dotted connectors denote presumed relationships suggested by participants or prior literature.}
    \Description{Four-column flow diagram summarizing how specific video elements and misalignment of cognitive characteristics, then to coping strategies, then to limitations  of coping strategies for BIF users in VBL. 
    Column 1 lists video elements: content requiring 2D to 3D translation; technical jargon and abstract terminology; high information density (visually dense scenes, long sentences, rapid pacing); single-channel information delivery; attention-intensive video; dialogue-embedded instructions; omitted scenes; and visual symbols used as inferential cues. 
    Curved arrows connect these items to Column 2, cognitive constraints: spatial perception limitations; verbal comprehension limitations; working-memory constraints (cognitive overload, processing speed, attention difficulties); and inferential-reasoning difficulties among concrete thinkers. 
    Arrows lead to Column 3, coping strategies: selectively seeking external support and repetitive viewing. 
    Arrows from these strategies point to Column 4, limitations: surface-level help-seeking only, mental and physical fatigue, and a knowledge barrier between video creators and BIF viewers.}
    \label{fig:result-summary}
\end{figure*}

This section examines participants' challenges and coping strategies in video-based learning, beginning with an example scenario illustrating how BIF users' cognitive characteristics manifest in practice.
We then explore how cognitive traits create specific video learning barriers (RQ1, Section~\ref{result-rq1}), experiential factors that amplify these challenges (RQ2, Section~\ref{result-rq2}), and the effectiveness of current coping strategies (RQ3, Section~\ref{result-rq3}), with Figure~\ref{fig:result-summary} summarizing the relationships between video elements, cognitive characteristics, and coping strategies observed in our study.

\subsection{A Video Learning Scenario of BIF Users}
We begin by walking through a participant's watching and learning session with instructional videos to better illustrate their observed challenges during the video-watching session.
P9, a 27-year-old male participant, watched a 2-minute AED instructional video produced by government while answering comprehension questions.
Reconstructed from researcher observations and participant responses, the following scenario demonstrates the cascading difficulties that emerged even with unlimited replay opportunities and researcher support.

P9 expresses confidence in understanding the content as the video is only about 2 minutes long.
As the video begins, he watches attentively as the narrator quickly explains AED basics while demonstrating the procedure.
The pace feels rapid, but he follows along as best he can.
P9 finds the pace challenging.
``\textit{It is definitely a bit complicated,}'' he says, though he tries to follow along as best he can.

Yet, he struggles to capture and retain the video content.
When trying to answer the AED pad placement quiz, P9 pauses and says ``\textit{I didn't see them (in the video)}.''
Similarly, when asked to answer true or false about continuing chest compressions while attaching pads, P9 struggles to recall details.
``\textit{I can't really remember},'' he says, requesting to view the video again.
After watching the relevant section again, he catches a subtitle mentioning compression continuity.
``\textit{It was subtitles},'' he explains with some satisfaction, marking his answer as correct.

The pattern of uncertainty continues as he moves through questions. 
When asked to choose the correct pad placement from four diagram options, P9 initially chooses option 2, the right answer.
When the researcher asks about his confidence, he responds ``\textit{Only 70\%.}''
But after rewatching the relevant section, confusion sets in.
``\textit{It was number 4},'' he revises, then immediately second-guesses himself, stating ``\textit{bit unsure between number 2 and number 4}.''
The researcher notes that despite multiple viewings, the angled camera perspective in the video makes spatial relationships difficult for P9 to discern clearly.

As the session progresses, P9's responses reveal concerning gaps in auditory processing.
When asked about voice guidance from the AED device, he confidently states, ``\textit{it didn't come out in the video}.''
Even after the researcher suggests listening more carefully for machine sounds, P9 cannot distinguish the AED's synthesized voice from the human narrator, despite both being present in the audio track.
The rapid pace and layered audio cues seem to overwhelm his processing capacity.

By the final questions about the defibrillation sequence, mental exhaustion becomes apparent.
``\textit{I lost track of it in my mind},'' P9 stated when asked about the steps following pad attachment.
The cumulative cognitive load of repeated viewing, sustained attention, and active recall has begun to take its toll.
Furthermore, throughout the session, even when P9 provides correct answers, he frequently expressed uncertainty about his responses.
This pattern of self-doubt persists despite having valid reasoning based on the video content.

\subsection{RQ1. Challenges Stemming from Misalignment between Cognitive Characteristics and Video Elements}\label{result-rq1}

In VBL, the challenges encountered by BIF users primarily stem from their individual cognitive characteristics, which become evident when interacting with video elements.
These cognitive traits manifested as observable behaviors when watching videos, revealing which aspects of video elements hindered their learning process.

\subsubsection{Spatial Perception Misalignment --- ``\textit{Is it the same, or is it a different machine?}''}

\begin{figure*}
    \centering
    \includegraphics[width=1\linewidth]{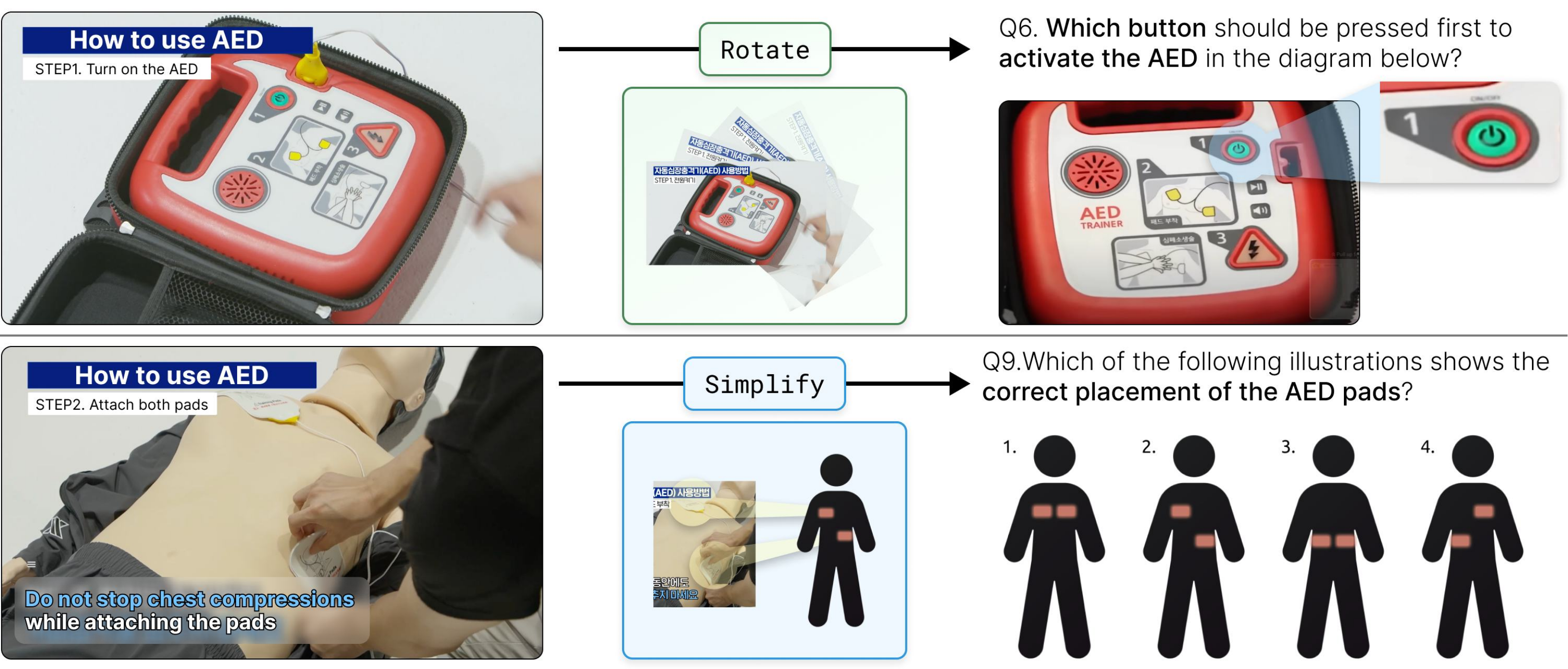}
    \caption{Examples of spatial perception and visual inference challenges in the AED instructional video. Top panel shows how camera angle and device rotation create spatial comprehension difficulties. Q6 tests whether participants can identify the power button when the AED is rotated 45 degrees from the video's viewing angle. Bottom panel (Q9) assesses pad placement comprehension, where participants must translate the angled view shown in the video to select the correct placement diagram among four options. These questions revealed significant spatial processing challenges for BIF users.}
    \Description{The figure illustrate how spatial transformations in instructional videos make it harder for BIF participants to comprehend. It was divided into two parts, upper and lower. 
    Top panels: On the left, the AED device is shown and rotated. On the right, a quiz question asks which button must be pressed first to activate the AED. The AED image appears differently due to the rotated viewing angle. 
    Bottom panels: On the left, a photograph shows pads being applied to a CPR mannequin. On the right, another quiz question asks participants to identify correct AED pad placement from four simple line drawings of a human figure with pads in different positions. Only one illustration matches the correct placement. This demonstrates the challenge of translating angled views in the video into correct understanding of pad positioning.}
    \label{fig:camera-angle}
\end{figure*}

During video-watching sessions, we observed diverse patterns of difficulty when participants attempted to comprehend the spatial relationships presented in the videos and transfer this understanding to reality.
\misc{
In Q6 (Appendix \ref{appendix-Q6}), participants had to recognize a rotated AED device and locate its power button (See Figure \ref{fig:camera-angle}, Top Panel) which was consistently shown in an angled view in the video.
P5 explicitly verbalized confusion, asking ``\textit{AED, so is it the same, or is it a different machine?}'' when encountering the rotated view.
Similarly, in Q9 (Appendix \ref{appendix-Q9}), several participants (P5, P6, P9, P10) struggled with pad placement shown from an angled perspective (See Figure \ref{fig:camera-angle}, Bottom Panel).
Notably, P9 initially selected the correct answer but confidently switched to an incorrect option after multiple viewings.}

\colormisc{These observations suggest that \textbf{participants faced spatial perception limitations impacting their ability to translate two-dimensional video demonstrations into three-dimensional physical actions.}
This challenge was particularly pronounced in replication-based how-to videos like cooking tutorials.
P7 described the frustrating gap between understanding and doing, mentioning ``\textit{I can understand it in my head, but when I actually try to do it myself, [...] my body just doesn't follow along.}''
SW1 corroborated this, noting ``\textit{they could explain it verbally 100\%. But when I asked them to actually take out a frying pan and do it, they could not.}''}

\colormisc{While translating graphics to 3D actions challenges all viewers \cite{tversky2002animation}, our findings reveal BIF users experience this challenge with significantly greater severity \cite{alloway2010working}.
This suggests that while this translating challenge affects all viewers, current video designs fail to provide the additional spatial scaffolding that BIF users require, creating a fundamental misalignment between content presentation and their cognitive capabilities.
}

\begin{figure*} [t]
    \centering
    \includegraphics[width=1\linewidth]{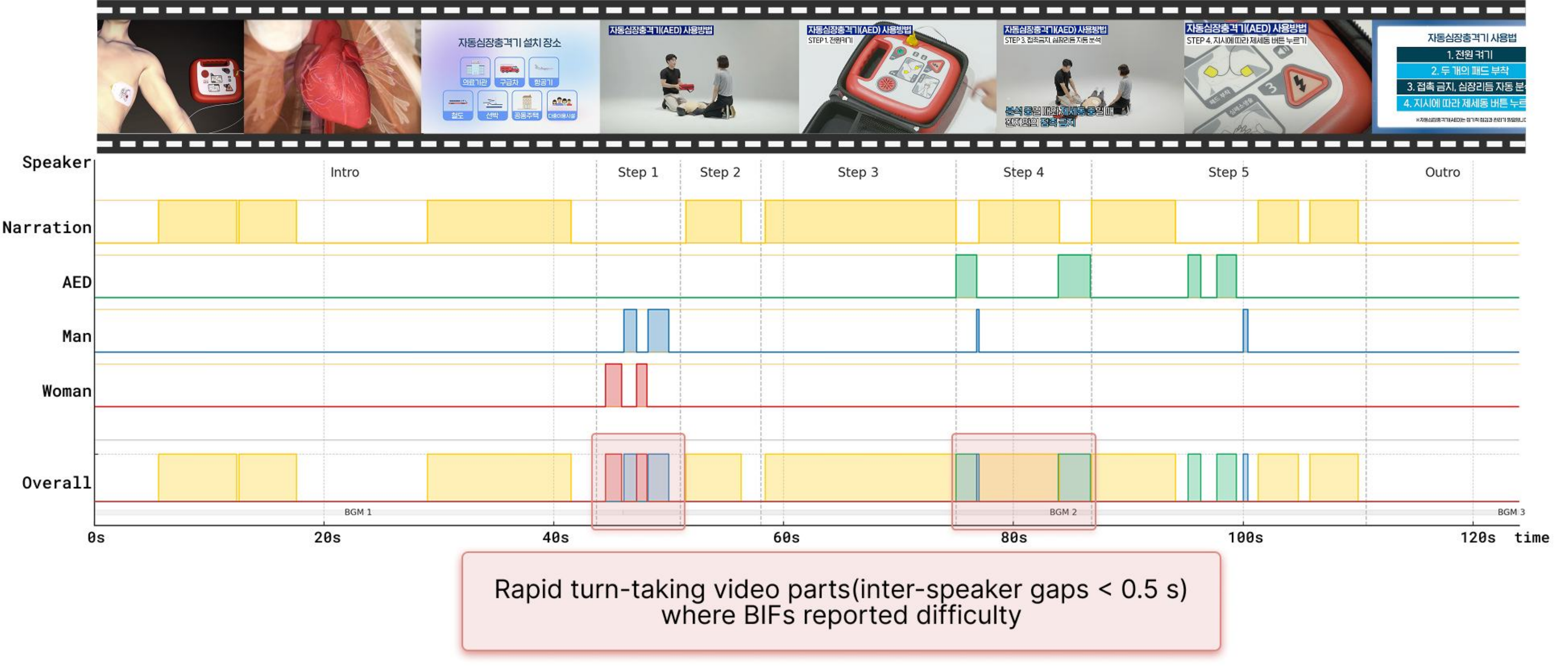}
    \caption{Timeline visualization of a 124-second AED instructional video showing parallel tracks for different speakers and content types. The video is segmented into sections (Intro, Steps 1-5, Outro) with tracks showing when the narrator speaks (yellow), AED device provides audio guidance (green), and dialogue occurs between a man (blue) and woman (red). Red boxes highlight periods of rapid turn-taking (inter-speaker gaps < 0.5 s) where BIF participants reported comprehension difficulties. Video stills above the timeline illustrate key moments from the instructional content.}
    \Description{
    The figure presents a timeline visualization of a 124-second AED instructional video. The horizontal axis shows time from 0 to 120 seconds, divided into labeled sections: Intro, Step 1, Step 2, Step 3, Step 4, Step 5, and Outro. Five parallel horizontal tracks display different audio/speaker elements: Narration (yellow blocks), AED device sounds (green blocks), Man speaking (blue blocks), Woman speaking (red blocks), and Overall combined audio (multicolored blocks at bottom). Red rectangular boxes highlight specific segments with rapid dialogue exchanges between speakers. Above the timeline, eight screenshot images from the video show scenes including AED device close-ups and people performing CPR on a mannequin. The visualization identifies moments where rapid turn-taking and overlapping audio create cognitive load challenges for BIF users.
  }
    \label{fig:overall-video}
\end{figure*}

\subsubsection{Verbal Comprehension Misalignment --- ``\textit{I don't understand what term is about.}''} 

Seven out of 12 participants exhibited markedly lower language comprehension thresholds, struggling with technical jargon and abstract terminology that neurotypical viewers would consider basic knowledge (P1, P2, P3, P6, P10, P11, P12).
The gap between creators' assumptions and users' actual comprehension became evident across various content types.

The comprehension gap persisted in how-to videos, where even supposedly `common sense' terminology created barriers.
P5 stated, ``\textit{even with computer knowledge that people think is common sense, understanding specific terminology about buttons, [...] such terms was the most difficult part.}''
These examples reveal that BIF users' language comprehension operates at a fundamentally different baseline, what content creators perceive as entry-level vocabulary often functions as technical jargon for this population.

\colormisc{
These findings suggest that \textbf{participants experienced a fundamental mismatch between their concrete thinking patterns and the abstract language in informational videos, creating comprehension barriers.}
Consequently, BIF users expressed strong preferences for simplified and concrete language.
For instance, P12 requested language accessible to ``\textit{elementary school students, even toddlers to understand}'' while P2 suggested interactive features where ``\textit{words light up}'' for immediate definitions.
}

These findings confirm established literature indicating that individuals with BIF are predominantly concrete thinkers who are most comfortable and competent when dealing with tangible, observable, and literal information~\cite{jankowska2012strategies, kim2024exploring}.
However, current video design practices fail to accommodate these linguistic differences, creating barriers to comprehension.

\subsubsection{Working Memory Misalignment}

Participants described challenges stemming from working memory constraints that manifested in three primary ways during VBL: \textbf{(i) cognitive overload from information density and rapid pacing, (ii) comprehension \colorfive{limitations} from single-channel information delivery, and (iii) concentration difficulties}.

\paragraph{Overwhelmed by Information Density and Rapid Pacing of Video --- ``\textit{It moved too fast, and I couldn’t keep up.}''} \label{results-overwhelmed}

During video-watching sessions, participants experienced severe working memory overload from rapid information delivery.
The 124-second AED video contained rapid turn-taking exchanges with pauses under 0.5 seconds (See Figure~\ref{fig:overall-video}), creating information density spikes that coincided with participants' expressed difficulties.
With the question describing situation of Step 1 (Q3, Appendix \ref{appendix-Q3}), four participants (P3, P7, P9, P11) showed fragmented comprehension and three (P2, P4, P5) reported complete inability to grasp content despite unlimited replays.

Participants identified three cognitive overload factors: (i) long sentences, (ii) high visual density, and (iii) rapid pacing.
% For instance, P11 noted, ``\textit{there were too many things on the screen, [...] and I get `blank'.}'' indicating not just temporary confusion, but complete cognitive overload where information processing halts.
\misc{While these also affect neurotypical viewers \cite{mayer2002multimedia, lang1999effects}, the threshold at which BIF users experience cognitive overload was significantly lower.
This mismatch between video pacing and BIF users' processing capacity \cite{alloway2010working} suggests they require more extensive content deconstruction to avoid the three cognitive overload factors, as SW1 emphasized, ``\textit{our standard is usually 2 to 3 times longer than teaching neurotypical children.}''
}

\paragraph{Limited Understanding by Single-Channel Information Delivery --- ``\textit{I feel like I need something like a caption or a highlight.}''}

\begin{figure*}
    \centering
    \includegraphics[width=1\linewidth]{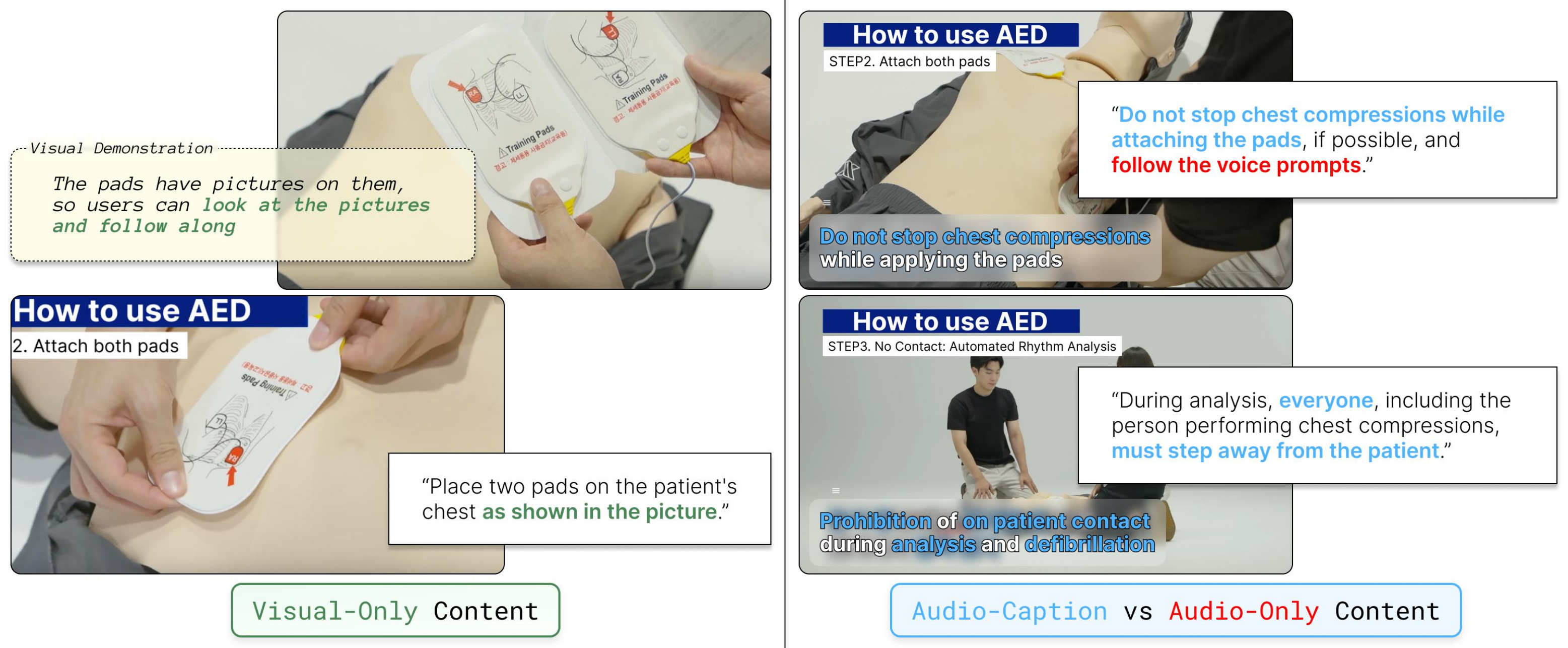}
    \caption{Examples of single-channel information delivery in instructional videos that create accessibility barriers for BIF users. Left panel shows "Visual-Only Content" where procedural information about pad placement is presented exclusively through visual demonstration without accompanying narration. Right panel contrasts "Audio-Caption" content (blue) where critical instructions are delivered with both narration and visual captions, versus "Audio-Only Content" (red) where the same instructions are delivered solely through narration without visual reinforcement. These examples demonstrate how BIF users struggle when information is presented through a single channel, highlighting their need for multi-modal presentation.}
    \Description{The figure is divided into two main panels demonstrating single-channel information delivery challenges in an AED instructional video.
    
    Left panel, titled "Visual-Only Content": Shows AED pad placement on a mannequin. The pads have printed diagrams indicating placement positions. The caption reads "Place two pads on the patient's chest as shown in the picture." Critical placement information is conveyed only visually through the demonstration, with no verbal explanation of specific positioning.
    
    Right panel, comparing delivery methods through color-coded texts:
    - Blue text labeled "Audio-Caption": Shows content where the instruction "Do not stop chest compressions while attaching the pads" is delivered with both narration and visual captions.
    - Red text labeled "Audio-Only": Shows where the instruction "Do not stop chest compressions while applying the pads, if possible, and follow the voice prompts" is delivered solely through narration without any visual reinforcement or captions.
    
    The color coding (blue for multi-modal, red for single-channel) visually emphasizes the contrast between accessible and inaccessible content delivery methods. Together, these panels illustrate how single-channel information delivery creates comprehension barriers for BIF users who require redundant multi-modal presentation to process complex procedural information effectively.}
    \label{fig:single-channel}
\end{figure*}

\misc{
Beyond pacing, we observed single-channel information processing \colorfive{difficulties} when content was presented exclusively through one modality (Fig.~\ref{fig:single-channel}).
When asked about AED voice guidance (Q11, Appendix \ref{appendix-Q11}) delivered solely through narration, five participants (P1, P3, P7, P9, P10) did not recognize that the AED provides voice prompts.
Similarly, seven participants (P1, P4, P5, P7, P9, P11, P12) misunderstood the blinking defibrillation button instructions (Q13-6) presented only through narration without caption.
Conversely, when information combined captions with narration (Q7-2, Q10, See Appendix \ref{appendix-Q7-2}), participants exhibited significantly higher comprehension (See Figure \ref{fig:single-channel}, Right Panel).
% This contrast reveals a critical insight that effective multi-modal delivery requires textual reinforcement of verbal content, not merely the presence of visual demonstrations alongside narration.
}

\colormisc{In addition, during interview, P3 requested visual emphasis for auditory information ``\textit{to emphasize that 'this is important right now',}'' while Parent4 noted ``\textit{visual and auditory elements combined would create much greater synergy.}''
These findings align with research showing BIF individuals' vulnerabilities in the phonological loop \cite{baddeley2020working}, explaining their audio-only comprehension \colorfive{difficulties} versus multi-channel success \cite{schuchardt2010working}, and visual feedback is more effective than auditory or tactile for people with cognitive disabilities \cite{kosch2016comparing}.
}

\paragraph{Concentration Difficulties --- ``\textit{Nothing enters my head. Whether I'm interested or not.}''}

Participants consistently reported difficulty maintaining sustained attention during video watching.
During the video-watching sessions, four participants (P5, P6, P9, P11) displayed nonverbal indicators of attention difficulties, including sighing and visible fatigue.
P5 explicitly connected prolonged viewing with physical discomfort, stating, ``\textit{if I watch it too much now, I'll feel sick.}''
This physical manifestation of cognitive fatigue suggests that attention difficulties for BIF users extend beyond simple distraction to encompass physiological responses to sustained mental effort.
\misc{
In addition, P5 explained that this difficulty remained constant regardless of topic interest, stating, ``\textit{whether I'm interested or not, I get tired first, and it doesn't enter my ears.}''}
 
Video length was identified as a critical factor in maintaining concentration.
\misc{
P1 and P12 established a specific threshold of 10 minutes.}
% , explaining \textit{``it seems difficult for me to concentrate for more than 10 minutes.''} (P12)
These findings confirmed previous research demonstrating that people with working memory \colorfive{constraints} are characterized by difficulties in concentrating \cite{martinussen2006working}, suggesting that the attention challenges observed in our study reflect the cognitive trait among BIF users.

\subsubsection{Inferential Reasoning Capability Misalignment}

While limited working memory and processing speed contribute to many comprehension barriers, another fundamental challenge for BIF users stems from difficulties with abstract thinking and inferential reasoning.
These cognitive processing differences manifest when videos require viewers to make connections between implicit information, interpret contextual cues, or bridge gaps in presented content.
We present three key inferential barriers observed during our study: \textbf{(i) challenges extracting actionable information from dialogue between characters, (ii) difficulty connecting omitted scenes and understanding sequential relationships, and (iii) difficulties utilizing visual symbols as interpretive cues}.

\paragraph{Difficulty Extracting Instructions from Character Dialogue --- ``\textit{I often have trouble understanding who the main character is, or who the villain is.}''}

\begin{figure*}
    \centering
    \includegraphics[width=1\linewidth]{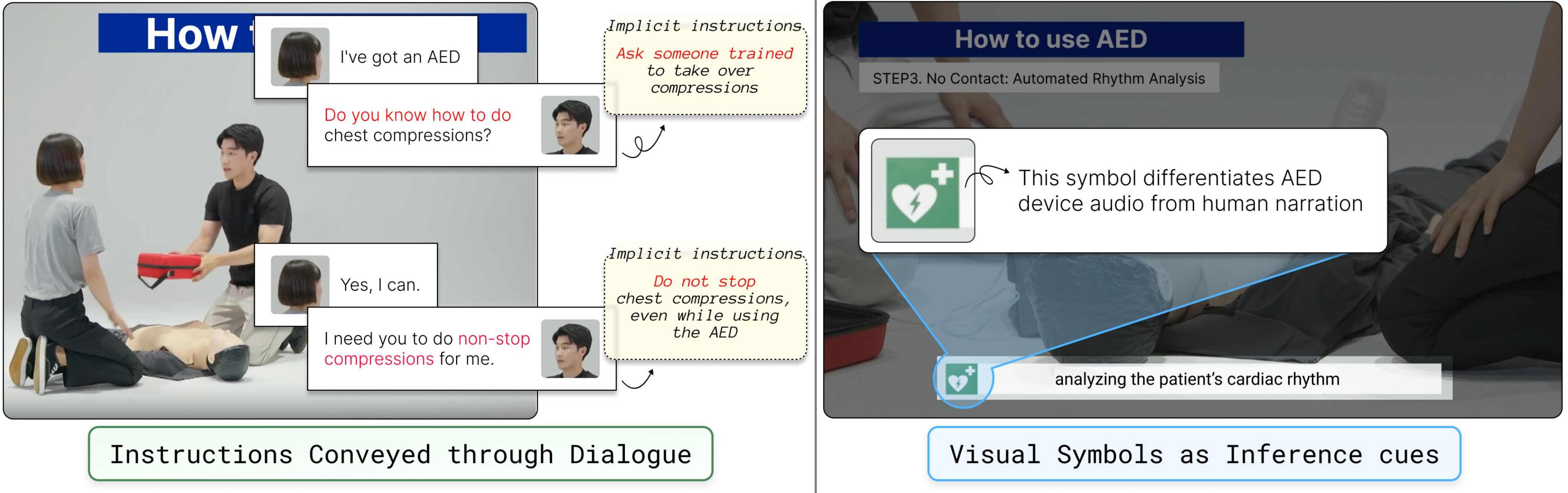}
    \caption{Examples of instructional content delivery methods in the AED video. Left panel shows ``Instructions Conveyed Through Dialogue'' where critical safety information is embedded within conversational exchanges between characters during an emergency scenario. Right panel shows ``Visual Symbols as Inference Cues'' demonstrating how the AED symbol is used in subtitles to distinguish machine-generated voice guidance from human narration, requiring viewers to make inferential connections between visual symbols and audio sources.}
    \Description{The figure is divided into two panels.
Left panel: A still image of two people practicing AED use is shown with dialogue overlaid. 
The dialogue reads: 
"I've got an automated external defibrillator."
"Do you know how to do (1) chest compressions?" 
"Yes, I can."
"I need you to do (2) non-stop compressions for me."  
Red highlighting emphasizes the instructions embedded in the conversation. 
A callout box summarizes the implicit safety rules: (1) Ask someone trained to take over compressions. (2) Do not stop compressions even while using the AED.  
This illustrates how instructions are indirectly conveyed through character exchanges.
Right panel: A video frame shows an AED training step. A green AED symbol with a heart and lightning bolt is highlighted. An annotation explains that this visual label is used to distinguish narration from sounds produced by the AED device. This demonstrates how viewers must infer meaning by connecting the visual symbol to the source of the audio.  
Together, the panels show two delivery methods in the instructional video, dialogue-based implicit teaching and symbol-based inferential cues.
  }
    \label{fig:delivery-methods}
\end{figure*}

A distinct comprehension barrier emerged when critical information was embedded within conversational exchanges rather than explicit instruction.
During the emergency scenario (Fig.~\ref{fig:delivery-methods}, Left Panel), viewers needed to extract safety protocols from dialogue between responders.
Beyond processing speed limitations (Section \ref{results-overwhelmed}), participants struggled to recognize that character interactions conveyed procedural instructions.
Among those who struggled with Q3 (P3, P7, P9, P11), some identified surface elements (speakers, CPR) yet missed the embedded safety protocol.
P9 heard ``\textit{didn't he ask for chest compression?}'' but could not determine why.
P5 similarly struggled, noting ``\textit{it's a give-and-take conversation [...] hard to explain.}''

This difficulty extended beyond instructional content.
P4 reported challenges with narrative media noting, ``\textit{I often find the drama's storyline very hard to understand. [...] I often have trouble understanding who the main character is, or who the villain is.}''
Parent1 observed similar patterns in their child's and stated ``\textit{they don't know how to do that (understand the flow from beginning to end).}''
Embedding critical information within character dialogue rather than direct instruction creates comprehension barriers across video domains.
This aligns with research demonstrating the inferential reasoning \colorfive{challenges} of individuals with BIFs, particularly when they must bridge between spoken dialogue and required actions \cite{kim2024exploring}.

\paragraph{Difficulty Connecting Omitted Scenes --- ``\textit{So we can't really figure out what the specific outcome is in the end.}''}

Informational videos frequently omit intermediate steps, requiring viewers to make bridging inferences about causal relationships. 
For BIF users with limited inferential reasoning, these gaps create comprehension barriers that repetition cannot overcome.
In the AED video, the transition from button press to CPR resumption omitted the actual electric shock delivery.
Eight participants (P2, P4, P6, P7, P8, P9, P10, P11) \colorfive{did not recognize} the connection between the ``step away'' warning with the button press (Q13-2, Q13-5, See Appendix \ref{appendix-Q13}).
Additionally, four participants (P2, P4, P7, P11) \colorfive{did not recognize} that pressing the button triggered the shock (Q13-4, See Appendix \ref{appendix-Q13}), with some believing the verbal warning itself delivered electricity automatically.
P6 exemplified this misinterpretation, stating, ``\textit{it said 'Step away from the patient.' I think people can receive an electric shock, just [by] an electric charge.}''

The video presented discrete elements—warning, stepping away, button press, CPR resumption—without explicitly connecting them.
P5 articulated the comprehension gap stating, ``\textit{we can't really figure out what the specific outcome is in the end.}''
Parent1 observed similar difficulties in narrative media noting ``\textit{while watching, they forget the beginning parts. They'd wonder, 'Why is that character over there?' the story doesn't connect.}''
This suggests that BIF users require explicit visual continuity and clear causal connections.
The cognitive demand of filling gaps while processing new information creates \colorfive{a misalignment with BIF users' working memory capacity}, resulting in fragmented understanding despite repeated viewing.

\paragraph{Difficulty Inferring Visual Symbols and Inferential Cues --- ``\textit{I didn't see that the green thing meant it was the AED.}''}

Participants described systematic difficulties utilizing visual symbols as inferential cues, even when they could recognize these symbols in isolation.
During the AED video watching, a critical challenge emerged in distinguishing audio sources.
The video employed an AED icon in subtitles to differentiate machine-generated voice guidance from human narration (Fig.~\ref{fig:delivery-methods}, Right Panel).
Despite this visual cue, seven out of 12 participants (P4, P5, P6, P7, P9, P10, P12) \colorfive{could not identify} which voice originated from the device (Q12).
P9 and P10 confidently asserted the AED produced no sound, while P12 stated``\textit{I thought the sound from the machine was the narrator's voice.}''

However, connecting the subtitle icon to its audio source, comprehension collapsed.
P7's mentioned this disconnection stating ``\textit{I just thought it was a regular subtitle... (Even I saw it) I didn't see that the green thing meant it was the AED.}''
The pattern revealed a fundamental cognitive barrier, while participants could perceive and identify symbols when explicitly prompted, they struggled to spontaneously utilize these same symbols as contextual cues during active viewing.
This finding confirms that established cognitive challenges with abstract reasoning and symbolic interpretation among BIF users create specific accessibility barriers in VBL contexts \cite{jankowska2012strategies}.

\subsection{RQ2. Experiential Factors Amplifying VBL Challenges}\label{result-rq2}
This section examines the challenges stemming from accumulated life experiences as individuals with BIF.
While these challenges are not solely stemming from VBL situations, the cognitive traits and accumulated life experiences of BIF users have led them to develop additional barriers that further complicate their engagement with informational videos.
% Aligning with previous literature \cite{kim2024exploring}, BIF users have developed low self-efficacy which works as a significant hurdle for VBL.
These experiential factors such as systemic 
exclusion to internalized self-doubt create psychological barriers that amplify the cognitive challenges identified in RQ1.

\subsubsection{\pointtwo{Systemic Exclusion and the Retreat to Video --- ``\textit{There aren't many people who can wait for us.}''}}
\colortwo{As individuals with BIF, they occupy a gray area \cite{peltopuro2014borderline} where they are cognitively different enough to struggle in neurotypical settings, yet excluded from specialized support systems offered for individuals with diagnosed IDD \cite{greenspan2017borderline, peltopuro2020borderline}.
This systemic exclusion manifests as chronic social pressure and interpersonal strain.
P7 stated, ``\textit{if people give us more time, we can probably do it. But society isn't like that, [...] there aren't many people who can wait for us, that's a problem.}''
The \textbf{constant pressure to keep up} with neurotypical pace creates an environment without adequate accommodation or recognition of their needs.
Faced with repeated implicit or explicit criticism, many individuals with BIF develop avoidance patterns as a coping mechanism \cite{hassiotis2008psychiatric}.
Parent1 observed this withdrawal in their child, mentioning ``\textit{they are very active and do well in online, but in reality, they have no people around them.}''
}

\colortwo{In this context, for individuals with BIF, VBL emerges not as an empowering educational choice, but as a \textbf{refuge from an unwelcoming social world}.
P4 articulated this motivation when explaining their preference for online lectures, stating ``\textit{I feel more comfortable. Because I don't have to face people directly, but I don't understand (online lectures) well (compare to in-person class).}''
This describes a painful paradox.
Videos offer psychological safety from social judgment, yet fail to provide adequate comprehension support.
BIF users retreat to VBL to escape the social anxiety and pressure of face-to-face learning environments, but they bring with them the accumulated trauma of exclusion, manifesting as diminished confidence and heightened anxiety about failure.}

\subsubsection{\pointtwo{The Trap of Uneven Abilities --- ``\textit{I have difficulty with numbers... but in humanities subjects, I perform above normal.}''}}

\colortwo{
Our findings reveal that the defining characteristic of the BIF experience is a \textbf{jagged cognitive profile} with significant intra-individual variance rather than uniform capacity reduction \cite{salvador2013borderline}.
Participants demonstrated awareness of these fluctuations, with P12 noting, ``\textit{I have difficulty with numbers... but in humanities subjects, I perform above normal.}''
Despite some universal barriers such as inferential reasoning difficulties, substantial variations in verbal and spatial skills led participants to \textbf{strategically leverage stronger cognitive domains to compensate for weaker ones.}
For instance, in the video-watching session, P2 prioritized auditory information to mitigate uncertainty in text processing, explaining, ``\textit{I listened first, and then looked at the subtitles [only] to verify if what I heard was correct.}''}

\colortwo{
However, this inconsistency creates a social trap of \textbf{misaligned expectations}, where competence in one area leads observers to misinterpret struggles elsewhere as a lack of effort.
The seeming success of compensatory efforts paradoxically masks underlying difficulties, resulting in ``invisible'' struggles and negative feedback from support networks.
SW1 highlighted how this invisibility leads to criticism, explaining ``\textit{due to characteristics where some parts work and some don't, it's difficult for parents to understand [...] (therefore) children receive a lot of negative feedback.}''
P4 recalled the pain of such misunderstanding, stating ``\textit{my parents were really frustrated, saying `Why can't you do something so easy?'}''
Consequently, this heterogeneity challenges accessibility approaches targeting specific barriers such as attention and reading in ADHD or dyslexia, necessitating personalization.
}

\subsubsection{\pointtwo{Negative Feedback Cycle and Internalized Sense of Inadequacy --- ``\textit{While `ordinary' people can do it at that speed, I'm quite lacking and slow in that.}''}}

\colortwo{Participants reported that accumulated learning gaps and continuous negative feedback have led to depression and low self-efficacy, as P8 noted feeling ``\textit{depressed}'' due to being ``\textit{slow.}''
This diminished confidence persisted during the study, manifesting as uncertainty even when participants exhibited accurate comprehension.
For instance, P11 and P2 rated their confidence at only 60\% and 43\% respectively despite providing correct answers.
P11 expressed his uncertainty, explaining, ``\textit{I think I remember seeing it in another video, but the expression in the question felt confusing.}''
Notably, P2 maintained this low confidence even after rewatching the video and confirming the subtitle stating ``\textit{don't stop compression while attaching pads,}'' revealing how deeply ingrained self-doubt overrides objective success.}

\colortwo{This self-doubt extended to a broader pattern where individuals with BIF \textbf{internalized universal challenges as personal failures}.
Participants consistently attributed difficulties, such as rapid video pacing, to their own limitations.
P1 exemplified this mindset, stating, ``\textit{while ordinary people can do it at that speed, I'm quite lacking and slow in that,}'' assuming that standard pacing is universally accessible.
However, this reveals a distorted self-perception, as prior research indicates that standard instructional video pacing is challenging for many viewers regardless of cognitive ability \cite{yang2022softvideo}.}

\subsection{RQ3. BIF Users' Current Coping Strategies for VBL Challenges}\label{result-rq3}

During VBL, participants employed three strategies when they encountered parts of the video they could not understand: (1) \textbf{Seeking External Support}, (2) \textbf{Repetitive Viewing}, and varied patterns of (3) \textbf{Speed Control of Video}.
% While repetitive viewing was universally adopted by participants and seeking external support was commonly attempted, speed control showed varied patterns of adoption.
However, our observations revealed that these strategies did not always guarantee successful problem solving and, in some cases, even led to new challenges, further undermining participants' confidence and sense of efficacy.

\subsubsection{\pointtwo{Seeking External Support --- Selective Help-Seeking. }}
\colortwo{
Our observation revealed participants' selective help-seeking behaviors.
While they were active in addressing technical barriers, they remained notably silent regarding content comprehension challenges.
When encountering technical obstacles such as screen sharing, nine out of 12 participants readily requested assistance from researchers (P1, P3, P4, P5, P6, P7, P9, P10, P11).
For instance, P3 asked ``\textit{what should I select here?}'' while P7 inquired ``\textit{how do I do screen sharing?}''
Similarly, P8 asked ``\textit{I think this is it, share screen. Yes, (then) what (do I have) to do?}''
These interactions indicate that participants' comfort with seeking help for execution-related challenges.}

\colortwo{However, this pattern contrasted with their silence regarding content comprehension difficulties.
Despite visible confusion during the video watching session, no participants asked clarifying questions about the video content itself.
Instead, they employed strategies such as (1) attributing to video factors, (2) guessing and seeking validation through the researchers' reactions.
For instance P4 attributed their confusion to external factors, stating ``\textit{I don't remember, the video went by too quickly.}''
In case of seeking validation through the researcher's reactions, P12 showed hesitation, mentioning ``\textit{I got confused [...] so B, no, it's not? Maybe...?}''}

\colortwo{The strategy that is closest to help-seeking for content-related factors was confirming their own interpretations.
For instance, P7 asked ``\textit{this AED, isn't that the red one?}''
Even when participants recognized their misunderstanding, they framed it as self-blame rather than a request for clarification, as P11 demonstrated by asking ``\textit{I misunderstood it as `to stop.' (It was)`do not stop'... am I being a bit stupid?}''}

\colortwo{This selective help-seeking pattern was consistent across participants.
Only two out of 12 participants reported their experience of requesting external help specifically for video content comprehension difficulties.
This number increased to only four participants when including those who had asked for help executing when something didn't work while watching how-to videos, but these requests still focused on technical functionality rather than content understanding.
Importantly, in independent home settings, where no ``researcher'' is available, this reluctance to seek content help suggests that confusion likely results not in help-seeking, but in disengagement or abandonment of the learning task entirely.}

\subsubsection{Repetitive Viewing as a Brute-Force Tool --- The Limits of Replaying. }
The majority of participants reported that they relied on repeated playback as their main coping mechanism in VBL and how-to videos, which extended from BIF users popular learning strategies --- rote memory \cite{jankowska2012strategies}.
While this approach sometimes proved effective, allowing participants to eventually grasp content through persistence and multiple viewings, it frequently fell short of addressing their comprehension needs.

The limitations of repetitive viewing became apparent in both persistent comprehension failures and consequent learning delays.
Many participants found that despite multiple attempts, certain content remained incomprehensible regardless of how many times they replayed it.
P2 explained, ``\textit{when I took online lectures, I watched one lecture about three or four times. [...] There were many things I could not understand even after repeatedly replaying them.}''

This created additional challenges, particularly the burden of delayed learning progress as participants spent disproportionate time on single concepts without advancing through the video. 
P1 mentioned, ``\textit{when I could not understand, I had to listen to the same part repeatedly, so I could not make much progress in my studying. That was also inconvenient.}''

This tendency was also observed in video-watching sessions.
Participants often mentioned watching repeatedly as the solution.
However, in video-watching sessions, even with an unlimited number of replays, several participants still struggled to answer correctly or experienced fatigue.
In video-watching sessions, nine out of 12 participants coped with repetitive playback, but several participants still could not arrive at correct answers even with unlimited replay opportunities.
Some participants even reported experiencing physical fatigue from the mental effort required during extended repeated viewing sessions.
For instance, P5 responded, ``\textit{if I watch it too much now, I don't feel well,}'' describing their physical discomfort.

Related to this limitation, participants wanted systems that could proactively manage their cognitive load by intelligently identifying and isolating content based on individual learning needs, rather than relying on manual replay functions.
P3 suggested, ``\textit{a feature that allows you to revisit specific points would be good. If I want to see important parts again, making a section repeat function so I can watch only that section again.}''

\subsubsection{Speed Control of Video --- The Efficiency-Understanding Paradox. }
Despite several participants reporting comprehension difficulties due to video pace, only two participants used slower playback speeds to enhance understanding (P6, P7).
The majority of participants instead kept the default playback speed (P2, P4, P5, P6, P7, P10, P11, P12), used speed-up functions (P3, P8, P9), or skipped unwanted portions of videos (P1, P8).
Participants explained that they avoided slowing down videos for several reasons.

This avoidance of slower speeds stemmed from multiple factors.
Participants reported that reduced playback speed created frustration, significantly extended viewing time, or they simply believed the default speed was already optimal.
P1 articulated this frustration, stating, ``\textit{if the speed is too slow, I tend to get frustrated, so I don't use it much.}''
Similarly, P11 expressed satisfaction with standard pacing, noting, ``\textit{I think the default speed is just right.}''

The paradox became particularly evident with participants who explicitly struggled with fast-paced content.
Despite these difficulties, they still refused to slow down playback.
For instance, P9 had mentioned difficulty understanding fast videos, yet when asked about using slower playback speeds, he responded, ``\textit{no, I haven't. I knew about it, but if you make it slower, it literally becomes slower and takes too much time, so I've never used it.}''
\misc{This resistance may also reflect an internalized pressure to maintain \textit{normalcy}.
Slowing down the video serves as a tangible admission of their ``slowness,'' triggering the feelings of inadequacy described in RQ2.
Thus, participants paradoxically cling to standard playback speeds to mirror neurotypical behavior, even at the cost of comprehension.}

This pattern persisted in video-watching sessions, where no participants used speed adjustment features for either fast-forwarding or slowing down.
Instead, several participants showed a tendency to skip sections they believed they had already understood, manually navigating through content rather than using built-in speed controls.
This preference for maintaining the default speed or increasing speed, despite acknowledging their own comprehension difficulties, reveals a concerning tension between participants' desire for efficiency and their need for deeper understanding.

\section{Discussion}
In our study, we examined the challenges, coping strategies, and expectations that individuals with BIF experience when learning from informational videos, and how their cognitive characteristics manifest as specific accessibility barriers.
We found that BIF users face multifaceted challenges stemming from misalignment of cognitive characteristics and video elements and experiential barriers that intensified VBL experience.
Despite employing coping strategies such as repetitive viewing, these approaches often prove insufficient for effective learning.
In this section, we synthesize our findings and propose design implications for BIF-inclusive video learning.
We then discuss how BIF users' experiences relate to broader cognitive accessibility challenges and address study limitations while outlining future research directions.

\subsection{\pointone{Understanding BIF Users' Video Learning Through Experiential, Cognitive, and Video Factors}}

\colorone{Our findings reveal that the challenges BIF users face in VBL stem not from isolated difficulties but from complex interactions between cognitive characteristics and experiential factors.
While previous research has established how cognitive domains interact, such as verbal comprehension supporting reasoning abilities \cite{kievit2017mutualistic, unsworth2014working}, our findings illuminate how these interconnections manifest as compounded barriers in video learning contexts when they are combined with BIF users’ unique social experiences.}

\colorone{\textbf{Experience of Negative Feedback and Retreat to VBL.}
Living in the gray zone between recognized disability and typical functioning, BIF individuals often lack formal accommodations available to other cognitively diverse populations \cite{wieland2016time}.
Consequently, they accumulate experiences of negative feedback in traditional educational settings, which often fosters an internalized deficit mindset.
Driven by a desire to ``mask'' their difficulties and avoid social stigma \cite{wieland2015psychopathology}, they retreat to VBL.
However, as challenges accumulate even in this self-directed environment, these difficulties are often attributed to personal limitations rather than design inadequacies.
This process progressively erodes self-efficacy and fosters \textit{learned helplessness} \cite{maier1976learned}, distinguishing BIF users.}

\colorone{\textbf{The Mismatch between Cognitive and Video Factors.}
However, this retreat leads to a new set of challenges caused by the friction between internal capabilities and external design.
Previous research has established how cognitive domains interact, for instance, verbal comprehension supporting reasoning abilities \cite{kievit2017mutualistic, unsworth2014working}.
Our study illuminates how these connections falter when BIF users' \textit{Cognitive Factors} (e.g., concrete thinking, limited working memory) collide with mainstream \textit{Video Factors} (e.g., rapid pacing, implicit semiotics).
While VBL has the potential to widen access \cite{mayer2002multimedia}, it paradoxically functions as a barrier when cognitive accessibility is overlooked \cite{BernabeCaro2020_TaxonomyE2U}.
For instance, the cognitive demand of translating 2D on-screen demonstrations \textit{(Video Factor)} into 3D physical actions overwhelms users with limited spatial perception \textit{(Cognitive Factor)}, turning a potential learning aid into a source of confusion.}

\colorone{\textbf{The Resulting VBL Deadlock.}
The convergence of these factors creates a ``learning deadlock.''
While the cognitive-video mismatch leads to initial comprehension failure, the experiential factors, specifically the internalized deficit mindset, paradoxically prevent users from utilizing effective coping tools or strategies.
Participants' reluctance to use speed controls, despite acknowledging pacing difficulties, reveals a critical disconnect between available tools and user needs.
Instead, they rely on ineffective strategies like brute-force repetition.
This cycle creates an invisible barrier where difficulties occupy a unique space: appearing surmountable yet consistently proving insurmountable without appropriate support.}

\subsection{Design Implications for BIF-Inclusive Video Learning}

\begin{table*}[t]
\centering
\caption{Design implications for BIF-inclusive video learning with content-level and Video UI-level improvements. The table presents nine design principles organized into three categories: Cognitive Load Reduction (addressing language complexity, visual density, and multi-modal balance), Scaffolding and Progressive Disclosure (focusing on visual markers, step-by-step guidance, adaptive segmentation, and personalization), and Fostering Self-Efficacy (emotional scaffolding, lowering help-seeking barriers). For each principle, we provide specific recommendations for both content creation and user interface design.}
\renewcommand{\arraystretch}{1.25}
\begin{tabular}{ 
    >{\centering\arraybackslash}p{2cm} | 
    >{\raggedright\arraybackslash}p{4cm} | 
    >{\raggedright\arraybackslash}p{4.5cm} | 
    >{\raggedright\arraybackslash}p{4.5cm} 
}
\toprule
\multicolumn{2}{c|}{\textbf{Design Implications}} & \textbf{Content-level Improvement} & \textbf{Video UI-level Improvement} \\
\midrule
\multirow{8}{*}{\textbf{\makecell[c]{Cognitive\\Load\\Reduction}}}
  & Use simpler language and provide interactive explanations
    & Replace jargon with plain language; match content to user's comprehension zone
    & Add clickable glossary terms; features where words light up for definitions \\
\cline{2-4}
  & Reduce visual density to improve clarity
    & Minimize extraneous visual elements; clear visual hierarchies for essentials
    & Provide a simplified/detailed view toggle; highlight key visuals with overlays \\
\cline{2-4}
  & Balance information across multiple modalities
    & Manage ``overall cognitive budget'' (e.g., decrease visual density if narration is complex)
    & Sync narration speed to visual density; provide adaptive pacing controls \\
\midrule
\multirow{11}{*}{\textbf{\makecell[c]{Scaffolding\\and\\ Progressive\\Disclosure}}}
  & Highlight important information with clear visual markers
    & Emphasize key objects/steps with consistent cues (colors, shapes)
    & Use structured overlays or animated highlights to draw attention \\
\cline{2-4}
  & Break down knowledge into step-by-step guidance
    & Make explicit steps; incorporate AI-generated explanatory scenes for omitted steps.
    & Provide causal continuity overlays between steps; interactive checkpoints \\
\cline{2-4}
  & Support adaptive video segmentation and strategic repetition
    & Divide content into conceptually distinct segments; pause at transitions
    & Auto-pauses at cognitive transition points; controls for targeted repetition \\
\cline{2-4}
  & \colorone{ Personalization for Strength-Based Modality Shifting}
    & \colorone{ Tailor dominant modality to user strengths (e.g., visual vs. auditory focus)}
    & \colorone{ Giving option for settings to minimize narration for visual cues or simplify audio descriptions} \\
\midrule
\multirow{5}{*}{\textbf{\makecell[c]{\colorone{ Fostering}\\ \colorone{ Self-Efficacy}}}}
  & \colorone{ Emotional scaffolding through automative pausing and non-evaluative check-ins}
    & \colorone{ Offer reassurance summaries; provide non-scored positive reinforcement}
    & \colorone{ Auto-pausing for validation check-ins; visual praise (e.g., filling progress bars)} \\
\cline{2-4}
  & \colorone{ Support non-verbal querying to  to lower help-seeking barriers}
    & \colorone{ Embed descriptive metadata for implicit cues and key objects}
    & \colorone{ Implement ``Click-to-Explain'' for objects/subtitles to reduce social stigma} \\
\bottomrule
\end{tabular}
\end{table*}

Based on our findings and participants' expressed needs, we propose the following core design implications that could significantly improve video accessibility for BIF users.
These principles are grounded in participants' actual requests, demonstrated needs, and current viewing strategies.

\subsubsection{Cognitive Load Reduction}

\textbf{Use Simpler Language and Provide Interactive Explanations. }
Our findings reveal that participants consistently struggled with jargon and abstract language, requesting more accessible alternatives with immediate definitional support.
This need manifests as both a comprehension barrier and a disruption to learning flow. 
As P2 suggested, an interactive approach like ``\textit{feature where just the words light up on the video and you can click them to know their meaning}'' would allow users to access definitions without disrupting the video flow or requiring external searches.
Videos can incorporate adaptive language translation that converts complex terminology based on BIF user's need, allowing users to choose their preferred comprehension level.
This is not merely about simplification but about matching content delivery to users' optimal comprehension zones.

\textbf{Reduce Visual Density to Improve Clarity.}
Our findings show a mismatch between the visual density of the video and the working memory capabilities of BIF users.
The cognitive overload participants experienced from visually cluttered scenes suggests a need for cleaner, more focused visual presentations.
Videos would become more inclusive by minimizing extraneous visual elements and providing clear visual hierarchies that guide attention to essential information \cite{kim2022fitvid}.

\textbf{Balance Information Across Multiple Modalities.}
Beyond managing language complexity and scene density, balancing information from multiple modalities to prevent cognitive overload of BIF users would be needed.
Cognitive overload stems not just from individual channels (language or visuals) but from the aggregate density across all modalities.
When complex language combines with dense visuals and rapid narration, video overflows BIF users' cognitive capacity.

Our findings suggest that video creators would benefit from considering the cumulative cognitive load across all channels rather than optimizing each in isolation.
One implication is an \textit{overall cognitive budget} approach, where increases in one modality’s complexity are balanced by simplification in others.
For instance, when presenting complex procedural steps, visual density could be decreased while narration slows, or when multiple visual elements are necessary, language could be simplified to provide cognitive space.

\subsubsection{Scaffolding and Progressive Disclosure}

\textbf{Highlight Important Information with Clear Visual Markers.}
Our co-watching sessions revealed that participants struggled to identify critical information when it was presented implicitly or embedded within complex visual contexts.
This challenge points to the need for explicit visual indicators that highlight important information, devices, or procedural steps. 
These markers would be beneficial if they are consistent and intuitive, helping users identify critical elements without competing for cognitive attention.
Using the consistent and structured colors, shapes, or animation styles to denote importance levels would help BIF users to understand marker itself, as this consistency reduces the cognitive load of interpreting new visual cues while helping BIF users develop pattern recognition for critical information.

\textbf{Break Down Knowledge into Step-by-Step Guidance to Reduce Inference.}
Our findings revealed a significant disconnect between creators' assumptions about viewer knowledge and BIF users' actual comprehension needs.
Creators often omit explanatory steps they consider obvious, while BIF participants struggle with these implicit transitions.
This gap manifests in multiple ways: creators assume familiarity with terminology, skip intermediate steps, and fail to explain the rationale behind actions.

Effective step-by-step support would be beneficial to make every procedural element explicit, provide rationale for each action, and include checkpoint moments where users can confirm their understanding.
Given participants' low self-efficacy, these checkpoints become crucial for ensuring accurate comprehension before moving forward. 

Furthermore, generative AI could be leveraged to address content gaps and inferential barriers.
First, it could automatically generate explanatory scenes for intermediate steps that creators may have assumed to be self-evident.
Second, it could provide \textit{causal continuity overlays} by rendering dynamic visual indicators, such as arrows or animated lines, that explicitly link cause-and-effect relationships between discontinuous scenes, helping BIF users maintain logical continuity.

\textbf{Support Adaptive Video Segmentation and Strategic Repetition.}
Participants expressed strong desires for more sophisticated content control that goes beyond basic replay functions.
To address this need, videos could automatically identify conceptually distinct segments and provide intuitive controls for targeted repetition.
Importantly, such repetition should be strategic, focusing on specific areas of difficulty rather than replaying entire sections wholesale.
A video interface can learn from user behavior to predict which segments require additional attention and automatically adjust pacing accordingly.
\pointone{However, given BIF users' limited ability to detect misinformation, such automated adaptations must prioritize accuracy and fail-safety.
Unlike neurotypical users who might spot an AI translation error, BIF users may internalize it as fact.
Thus, these systems function best as curated overlays rather than purely generative interventions, ensuring that adaptive support enhances rather than compromises content reliability.}

In addition, features such as controlling information density by providing breaks for processing could be helpful.
The overwhelming nature of rapid information presentation requires systematic approaches to pacing control.
Videos can incorporate automatic pauses at cognitive transition points, allowing users to process information before proceeding.
This addresses the fundamental mismatch between video pace and BIF users' processing speeds.

\pointone{
\textbf{Personalization for Strength-Based Modality Shifting. }
Recognizing that BIF users occupy a spectrum with diverse cognitive profiles, a one-size-fits-all approach is insufficient.
Strength-based personalization strategy that adapts content delivery to individual capabilities would be helpful.
Rather than simply simplifying all content, the system would be beneficial to allow for modality shifting based on user strengths.
For instance, a user with weaker auditory processing but strong visual recognition could receive content where narration is minimized in favor of enhanced visual cues and captions. 
Conversely, a user who struggles with reading could benefit from simplified audio descriptions.
By tailoring the dominant modality to the user's cognitive strengths, the system can reduce friction and maximize engagement without stripping away necessary information.
}

\subsubsection{\colorone{Fostering Self-Efficacy}}

\colorone{
Our findings indicate that BIF users often possess the correct knowledge but lack the confidence to assert it, exhibiting low self-efficacy.
To address this, we propose design interventions that shift the system's role from evaluation to reassurance and support.}

\colorone{\textbf{Emotional Scaffolding through Automative Pausing and Non-Evaluative Check-ins. }
Previous research indicates that frequent evaluation reinforces fear of failure, framing every interaction as a potential test \cite{elliot20012}.
Therefore, shifting from assessment to reassurance is essential for BIF users.
Instead of testing users with questions like ``\textit{What is X?}'', the system can provide non-evaluative check-ins, potentially triggered by automatic pausing at key moments, to validate progress (e.g., ``\textit{We just covered [Key Summary]. Ready to move on? It's okay to review.}'').
This approach can transform the interaction from a test of memory to a confirmation of engagement.
For instance, when users acknowledge their understanding, the system would be helpful to provide immediate, non-scored positive reinforcement (e.g., visual praise, progress bars filling up), validating their journey and building the confidence required for self-directed learning.}

\colorone{However, this reinforcement would benefit by distinguishing between effort and accuracy.
While the system validates the user's persistence (e.g., ``\textit{Great job sticking with this!}''), feedback on safety-critical comprehension (like AED use) must remain objectively accurate to prevent a false sense of competence.
The goal would be building confidence in learning capability, not masking comprehension gaps.
}

\colorone{\textbf{Support Non-Verbal Querying to Lower Help-Seeking Barriers. }
Help-seeking represents an adaptive learning strategy through which learners pursue successful learning independently.
Such help-seeking behaviors are categorized into instrumental help, where learners seek `methods' to solve problems themselves, and executive help, where learners seek `immediate solutions' from others \cite{nelson1985helpseeking}.
Our findings reveal that participants rarely engaged in instrumental help-seeking.
This pattern reflects how metacognitive judgment (assessing whether one can solve the problem independently), along with shame, self-esteem concerns, and fear of social evaluation, influence the execution of help-seeking behaviors \cite{karabenick2011understanding}. }

\colorone{To reduce the friction in reaching the point of asking questions and minimize social stigma, systems may benefit from supporting non-verbal querying mechanisms, such as \textit{click-to-explain}.
Alternatively, systems could present frequently asked questions by other learners, enabling users to select from pre-formulated queries rather than articulating their own.
Allowing users to bypass the formulation of complex questions can enable BIF users to resolve ambiguities immediately without the cognitive burden of linguistic formulation or social pressure.}

In summary, these design implications highlight the importance of reducing cognitive load and providing progressive scaffolding to create more accessible and effective video learning experiences for BIF users.
\colorone{Beyond these general principles, personalization based on more precise user modeling is also worth considering.}
Given the \textit{jagged} cognitive profiles of BIF individuals and their tendency to mask learning difficulties, this context underscores the potential usefulness of implicit, behavior-based data collection alongside explicit self-reports.
Different users may benefit from varying degrees of language simplification, pacing, or scaffolding depending on their knowledge and preferences.
For instance, terminology could be explained at an elementary level for some learners and in more advanced terms for others, while UI features such as adaptive segmentation or interactive explanations could adjust based on individual needs.
Such personalization would help these strategies remain flexible and responsive to diverse user profiles.
% 이런 것들이 되려면 결국 user modeling이 필요하고, robust한, precise한 User modeling이 결국 중요하고 잘 되면, personalization에 활용될 수 있다.

\subsection{\pointone{Situating BIF Challenges Within the Broader Cognitive Accessibility Landscape}}

% 얘네 이거 좋은데 얘네한테 적용되어도 좋을 것 같음. 다른 특징을가진 집단이기때문에 overlap되어도 되지 않을까

% \colorone{BIF users occupy a distinctive position in accessibility research, falling between neurotypical functioning and recognized disabilities.
% Comparing their experiences with other cognitively diverse populations reveals that while needs often overlap, the underlying mechanisms and necessary design interventions differ.}
BIF users occupy a distinctive position in accessibility research, falling between neurotypical functioning and recognized disabilities.
Comparing BIF users with other cognitively diverse populations shows that, although some accessibility needs appear similar, the underlying cognitive mechanisms and appropriate design interventions differ.

\colorone{First, while BIF users share attention-related challenges with individuals with ADHD \cite{jiang2025shifting}, the sources of these difficulties differ, leading to different design needs.
ADHD challenges primarily stem from executive dysfunction and attentional control, whereas BIF users face a compound burden of working memory constraints, slow processing speed, and limited inferential reasoning operating simultaneously.}

Similarly, while BIF users and individuals with dyslexia both experience difficulties when engaging with text-based content \cite{mccarthy2010dyslexia}, the nature of these difficulties is different.
BIF users struggle with abstract concept comprehension and inferential reasoning even when decoding is successful.
This suggests that text simplification or attention management tools which are common solutions for Dyslexia and ADHD are insufficient for BIF users who require deeper scaffolding to bridge abstract concepts and concrete understanding.

\colorone{Furthermore, while BIF users and individuals with IDD share challenges related to abstract reasoning and verbal working memory \cite{esposito2024advancing, l2025accessible}, they differ in how support is mediated.
Unlike many IDD populations, who often have access to caregivers that mediate accessibility barriers, our findings show that BIF users typically navigate unmodified mainstream environments (e.g., standard YouTube interfaces) without external support, remaining under-recognized by support systems.

Our study addresses this gap by proposing BIF-specific design guidelines that prioritize independent adoptability.
In contrast to IDD-focused designs that assume caregiver assistance for setup and configuration, tools for BIF users would benefit from being grounded in users’ existing video-viewing practices, reducing the cognitive burden of configuration complexity, and embedding intrinsic scaffolding that remains effective without caregiver intervention.}

\colorone{Finally, while BIF users and older adults share challenges related to cognitive fatigue and reduced processing speed \cite{kim2023older}, they differ in foundational verbal comprehension and conceptual understanding.
Older adults typically retain strong verbal comprehension and vocabulary despite declines in processing speed, allowing them to grasp abstract concepts given sufficient time \cite{verhaeghen2003aging, park2009adaptive}.
In contrast, BIF users face fundamental challenges with abstract reasoning and technical terminology, meaning that simply slowing down the video is insufficient without additional scaffolding to bridge the conceptual gap.}

\colorone{Across these comparisons, BIF users share observable challenges with other populations, while differing in the cognitive and social mechanisms that shape how these challenges arise and how they need to be addressed.
Accordingly, BIF users may benefit from design approaches that are specifically tailored to their profile, rather than direct adaptations of frameworks developed for ADHD, IDD, or aging populations.
Their status as a ``borderline'' population that seeks independence while facing cognitive processing challenges in standard interfaces demands adaptive systems that recognize their specific profile: functional enough to access mainstream media, yet cognitively vulnerable enough to fail without invisible, integrated support.}

\subsection{Limitations and Future Work}

Several limitations of our study need to be acknowledged.
First, as our sample of 12 BIF participants aged 23-33 represents only young adults and missing BIF users in other age groups whose video learning needs and strategies may differ.
We also interviewed parents who are representatives of regional BIF caregiver support groups in South Korea's public communities, not our participants' own parents, potentially missing individual-specific insights.

Second, our observational study focused on a single topic video. While it provided valuable insights, it limited our exploration on how BIF users engage with diverse content types across domains.
Future investigations that study the combined effects of topic complexity and diversity would be needed to develop more comprehensive accessibility solutions.

% Third, our single-session study with fixed topic video, while providing valuable insights, limited our exploration on how BIF users engage with diverse content types across domains.
% Future research would be helpful to examine different videos varying in complexity, and production quality.

Last, our quiz-based observational study, while revealing real-time comprehension challenges, was conducted in a controlled research environment that may not fully reflect natural video consumption patterns.
Longitudinal studies examining BIF users' video learning behaviors in naturalistic settings would provide valuable complementary insights.

Future research can also explore how strength-based design principles can be implemented and evaluated in real-world video learning contexts, particularly examining the long-term learning outcomes of such approaches.
In addition, research into automated systems for content adaptation, such as AI-powered language simplification or intelligent content segmentation, could make these accessibility improvements scalable across vast video libraries.
Finally, the development of assessment tools for evaluating video accessibility from a cognitive perspective would also advance the field, providing creators and platforms with concrete metrics for measuring and improving inclusivity.

\section{Conclusion}

\misc{As video-based learning (VBL) becomes essential for acquiring practical skills, especially those with cognitive differences, is critical.
This study aimed to deeply understand the accessibility challenges faced by individuals with Borderline Intellectual Functioning (BIF) in VBL. 
We found that that current video designs create unique barriers through rapid pacing and implicit information, which, combined with experiential factors like low self-efficacy and ineffective coping strategies, significantly hinder learning efficiency.
By identifying these specific misalignments, we propose targeted design implications that not only guide inclusive content creation but also inform the development of intelligent adaptation tools.
Ultimately, this work aims to eliminate the obstacles faced by individuals with BIF when utilizing the online video resources for learning, enabling BIF users to leverage digital learning resources with greater confidence and independence.}

\begin{acks}
This work was supported by the National Research Foundation of Korea (NRF) grant funded by the Korea government (MSIT) (No.RS-2024-00406715)
This work was also supported by Institute of Information \& Communications Technology Planning \& Evaluation (IITP) grant funded by the Korea government (MSIT) (No.2021-0-01347,Video Interaction Technologies Using Object-Oriented Video Modeling).

We extend our gratitude to all participants who generously shared their experiences with video-based learning and invaluable insights that shaped this research. Special thanks to our reviewers for their constructive feedback that strengthened this work, and to KIXLAB members (especially Yoonseo) for their continued support throughout this journey.
\end{acks}

% \bibliographystyle{ACM-Reference-Format}
\bibliographystyle{ACM-Reference-Format}
\bibliography{reference}

\clearpage
\appendix

% Going to add content about how we developed probe
\section{Appendix}

\subsection{Interview Questionnaire of Interviews with Social Workers and Parents with BIF Children} \label{appendix-SWParentInterview}
\subsubsection*{1. Warm-up}
\begin{itemize}
    \item Introduction and appreciation:  
    Thank you for participating in today’s interview. My name is [Researcher’s Name]. This interview is part of a study on developing how-to videos to support students with borderline intellectual functioning.
    \item Purpose of the interview:
    We aim to gain insights, based on your experiences and perspectives, that will help us design video content useful in practice.
    \item Process of the interview: 
    The interview will be guided by prepared questions, but you are welcome to share any additional thoughts at any time.
    \item Consent and recording:
    The content will only be used for research purposes. With your permission, we would like to record the interview for analysis. Would that be acceptable?
    \item Participant introduction:
    Could you briefly introduce yourself?  
    Could you describe your current role as a social worker?  
    What kinds of experiences have you had working with students with borderline intellectual functioning?
    \item Perceptions of borderline intellectual functioning:
    In your view, what are the key characteristics or definitions of borderline intellectual functioning?  
    How do you perceive their general learning and behavioral characteristics?
\end{itemize}

\subsubsection*{2. Learning Characteristics of BIF individuals}
\begin{itemize}
    \item What are the major challenges or difficulties you observe in BIF individuals’ learning processes?  
    \item What difficulties do BIF individuals usually experience when learning through videos? What kinds of support do you think are most necessary?
    \item In what situations do BIF individuals experience particular difficulties with video-based learning?  
    \item What types of support would you consider most effective in addressing these challenges?
\end{itemize}

\subsubsection*{3. Opinions on Existing Support Methods and Tools}
\begin{itemize}
    \item What strategies or support methods do you currently use for BIF individuals (e.g., counseling, group activities, digital tools)?  
    \item What are the advantages and disadvantages of video or multimedia resources in this context?  
    \item If you have experience using such tools, how effective do you consider them?  
    \item Which aspects of these methods seem effective, and which require improvement?
\end{itemize}

\subsubsection*{4. Perceptions of Video-Based Learning}
\begin{itemize}
    \item What do you think are the positive effects of video-based resources on BIF individuals’ learning?  
    \item What limitations or concerns do you see regarding the use of such tools?  
    \item Based on your experience, what were the most impressive or disappointing aspects of using video resources?
\end{itemize}

\subsubsection*{5. Ideas and Requirements for Future Development}
\begin{itemize}
    \item If you have previously used similar video resources, what aspects were most helpful?  
    \item If not, what elements do you think should be included in effective informational videos?  
    \item In your opinion, would approaches such as storytelling, case-based learning, or interactive elements be effective?  
    \item What should be considered to make video content more engaging and accessible for BIF individuals?  
    \item From a user interface and accessibility perspective, what features should be prioritized?
\end{itemize}

\subsubsection*{6. Feedback and Recommendations for Development}
\begin{itemize}
    \item Could you share a success story or, conversely, a case that highlighted areas for improvement?  
    \item What specific advice would you give for the development of informational videos?
\end{itemize}

\subsubsection*{7. Closing and Additional Comments}
\begin{itemize}
    \item Do you have any additional thoughts or suggestions regarding the development process or research?  
    \item Thank you for sharing your valuable insights today. Please feel free to add any final remarks or questions.
\end{itemize}

\subsection{Interview Questionnaire of Study with BIF Users} \label{appendix-BIFInterview}
\subsubsection*{1. Warm-up}
\begin{itemize}
    \item Introduction and appreciation: 
    Thank you for joining today’s interview. My name is [Researcher’s Name]. In this session, we would like to hear about your experiences and challenges when learning through videos such as YouTube or school classes. Please feel free to share your thoughts comfortably.  
    \item Purpose of the interview:
    We aim to gain insights from your experiences to develop a system that makes tutorial and educational videos (e.g., on YouTube) easier to understand and use.  
    \item Process of the interview:
    The interview will be guided by prepared questions, but you can skip any questions you do not wish to answer. If you prefer not to speak, you may also use the Zoom chat to share your response.  
    \item Consent and recording:
    For anonymity, you may choose a nickname or preferred form of address. The interview will be audio-recorded (camera use is optional) and used only for research purposes with your consent. Would this be acceptable?  
    \item Participant introduction:
    Could you briefly introduce yourself?  
    Could you also describe your learning environment at home or school, and your daily role as a learner?  
    \item Experiences in standardized learning contexts:
    Have you experienced difficulties following standardized learning settings such as school or academy classes? Could you describe those situations?  
    \item Experiences in non-standardized learning contexts:
    Have you experienced difficulties when learning new skills or tasks at work or part-time jobs? If so, could you share specific examples?  
    \item Cognitive strengths and weaknesses: 
    Based on past assessments (e.g., Wechsler scales), are there particular areas where you face the most challenges (e.g., verbal comprehension, working memory, perceptual reasoning, processing speed), or do you experience difficulties more broadly?  
\end{itemize}

\subsubsection*{2. Video Viewing Habits and Experiences}
\begin{itemize}
    \item How often do you usually watch videos? (e.g., times per week, average hours per day)  
    \item What types of videos do you usually watch (e.g., entertainment, learning, hobbies)?  
    \item Do you watch videos to learn new things? If yes, what kinds of topics have you learned (e.g., cooking, online lectures, certification prep)?  
    \item How do you usually search for videos that suit your needs?  
    \item Do you use additional tools such as playback speed control, subtitles, or AI tools when watching videos? Were these helpful?  
    \item Which kinds or video were most useful or memorable?  
    \item When selecting informational videos, what criteria are most important to you?  
    \item Could you share a specific example of a helpful informational video and one that was difficult to follow?  
\end{itemize}

\subsubsection*{3. Difficulties in Video-Based Learning}
\begin{itemize}
    \item Concentration: Have you experienced difficulties focusing on videos (e.g., getting distracted, struggling to find the right part again, long video length, too many elements on screen, background noise)?  
    \item Motivation: Have you ever stopped watching because the video felt boring, lacked rewards, or you could not see the reason to continue?  
    \item Social context: Have you found it difficult to understand jokes or social cues in tutorial videos, and did that affect your learning?  
    \item Comprehension: Have you struggled with understanding words, explanations, or complex content in videos?  
    \item Searching: Have you had difficulties finding appropriate videos or information within a video?  
    \item Interface and functions: Have you experienced challenges when trying to use playback speed, subtitles, or other assistive features? Were they helpful or inconvenient?  
    \item Other difficulties: Are there any other barriers you would like to share?  
\end{itemize}

\subsubsection*{4. Ideas and Requirements for Future How-to Video Development}
\begin{itemize}
    \item What are the main difficulties you personally experience when learning through videos?  
    \item What kinds of support would help you most?
    \item Have you ever used AI-based features such as automatic video summarization? If yes, what was helpful or insufficient?  
    \item What improvements or additional functions would make video-based learning easier for you?  
    \item What should be considered to make video content more understandable for BIF individuals?  
    \item From a user interface and accessibility perspective, what features (e.g., subtitles, playback control, assistive tools) are most important?  
    \item Do you have concerns about technical or usability limitations of video-based learning?  
    \item Based on your experiences, what parts of video-based learning should be improved, and what ideas would you suggest?  
\end{itemize}

\subsubsection*{5. Closing and Additional Comments}
\begin{itemize}
    \item Do you have any additional thoughts or suggestions related this research?  
    \item Thank you very much for sharing your experiences and thoughts today. Your input will help us find ways to improve video-based learning for individuals with BIF. Please feel free to contact us anytime if you wish to add further comments.  
\end{itemize}

\subsection{Co-watching Session AED Questionnaire}\label{appendix-quiz}

\pointfour{The comprehension quiz was designed to probe multiple dimensions of video understanding that could reveal accessibility barriers for individuals with BIF. Rather than assessing isolated cognitive abilities, questions were structured to reflect the integrated nature of video comprehension, where understanding often requires simultaneous processing of visual, auditory, and procedural information. Questions were organized by video segments (Intro, Steps 1--4, Outro) and designed to capture difficulties that might not emerge through self-report alone. Below, we provide the rationale for each question followed by the questions themselves.}

\textit{(The correct answer is shown in bold.)}

\subsubsection*{Intro}

\noindent\textbf{Q1.} Which of the following statements about an Automated External Defibrillator (AED) is correct?
\begin{itemize}
  \item (A) An AED is a device that injects medicine into a person whose heart has stopped.
  \item (B) An AED is a device that raises the body temperature of a person whose heart has stopped.
  \item \textbf{(C) An AED is a device that delivers an electric shock to restart the heart of a person whose heart has stopped.}
  \item (D) An AED is a device that measures the blood pressure of a person whose heart has stopped.
\end{itemize}

\pointfour{\noindent\textit{Rationale: The correct answer uses nearly identical wording to the video narration. This tests recognition and retention of audio-only content.}}

\vspace{0.5em}
\noindent\textbf{Q2.} Answer O/X (True/False). An AED can only be used in hospitals, ambulances, airplanes, trains, large buildings, and crowded places. (O/\textbf{X})

\pointfour{\noindent\textit{Rationale: Tests retention of information delivered through both audio and visual channels. We also observed whether participants attended to the word ``only'' when reading the question, rephrasing verbally when needed to assess video comprehension rather than text reading ability.}}

\subsubsection*{Step 1}

\begin{figure}[H]
    \centering
    \includegraphics[width=0.8\linewidth]{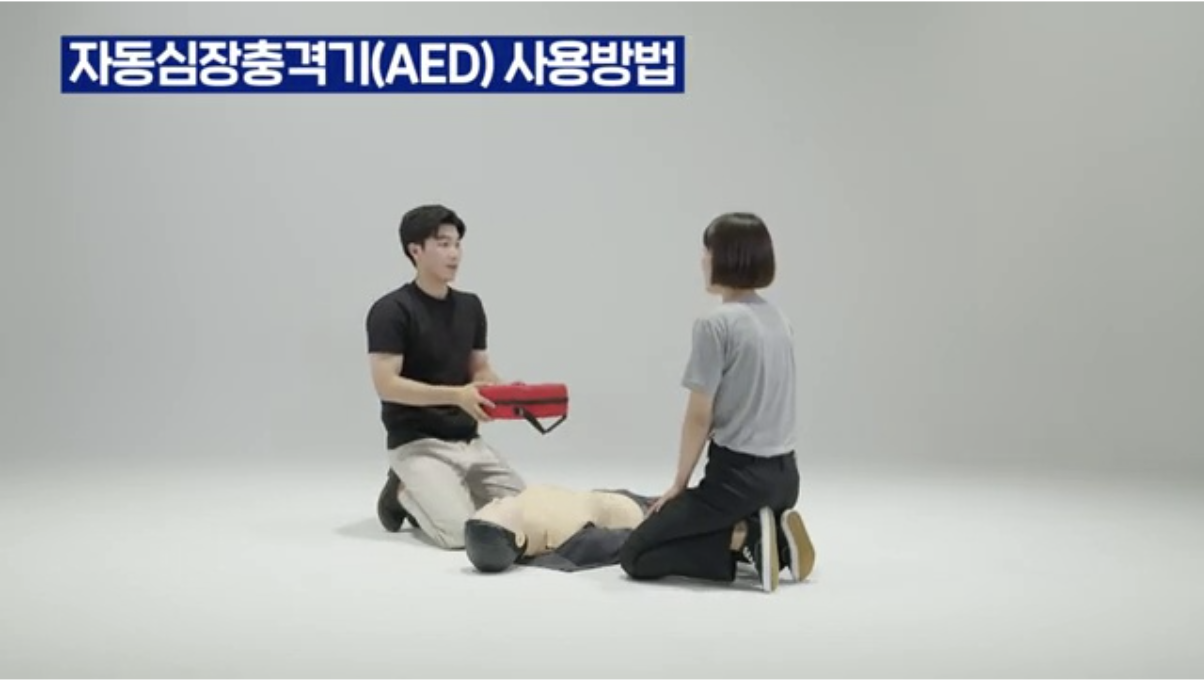}
\end{figure}
\noindent\textbf{Q3.} Please watch the video again and explain the situation in the video. What are the roles of the man, woman, and lying mannequin? \label{appendix-Q3}

\textbf{Model Answer:}
\begin{itemize}
  \item \textbf{Man:} Performing chest compressions (CPR) on the collapsed person.
  \item \textbf{Woman:} Handing over the red bag (AED).
  \item \textbf{Mannequin lying down:} The patient in need of help.
\end{itemize}

\pointfour{\noindent\textit{Rationale: Tests ability to identify roles when key information is conveyed implicitly through dialogue and actions rather than explicit narration.}}

\vspace{0.5em}
\noindent\textbf{Q4.} In the video, what request did the person performing CPR make to the one who went to get the AED?
\begin{itemize}
  \item (A) Asked to stop CPR and bring the AED.
  \item \textbf{(B) Asked to continue CPR without stopping.}
  \item (C) Asked to press harder during CPR.
  \item (D) Made no request.
\end{itemize}

\pointfour{\noindent\textit{Rationale: Tests ability to extract key instructions from continuous dialogue.}}

\vspace{0.5em}
\noindent\textbf{Q5.} If, in a situation like in the video, your partner says they don't know how to perform chest compressions, what is the best action?
\begin{itemize}
  \item \textbf{(A) I perform chest compressions and ask the partner to prepare the AED.}
  \item (B) I explain how to do chest compressions and ask the partner to continue compressions without stopping in my place.
  \item (C) Give up using the AED and continue chest compressions as before.
  \item (D) Pause chest compressions temporarily and prepare the AED.
\end{itemize}

\pointfour{\noindent\textit{Rationale: The video implicitly conveys the importance of continuous chest compressions through dialogue. This tests whether participants recognized this information (from prior knowledge or the video) and can apply it to a real-world scenario. Follow-up questions probed their reasoning process.}}

\subsubsection*{Step 2}

\noindent\textbf{Q6.} Which button should be pressed first to activate the AED in the picture below? \label{appendix-Q6}
\begin{figure}[H]
    \centering
    \includegraphics[width=0.8\linewidth]{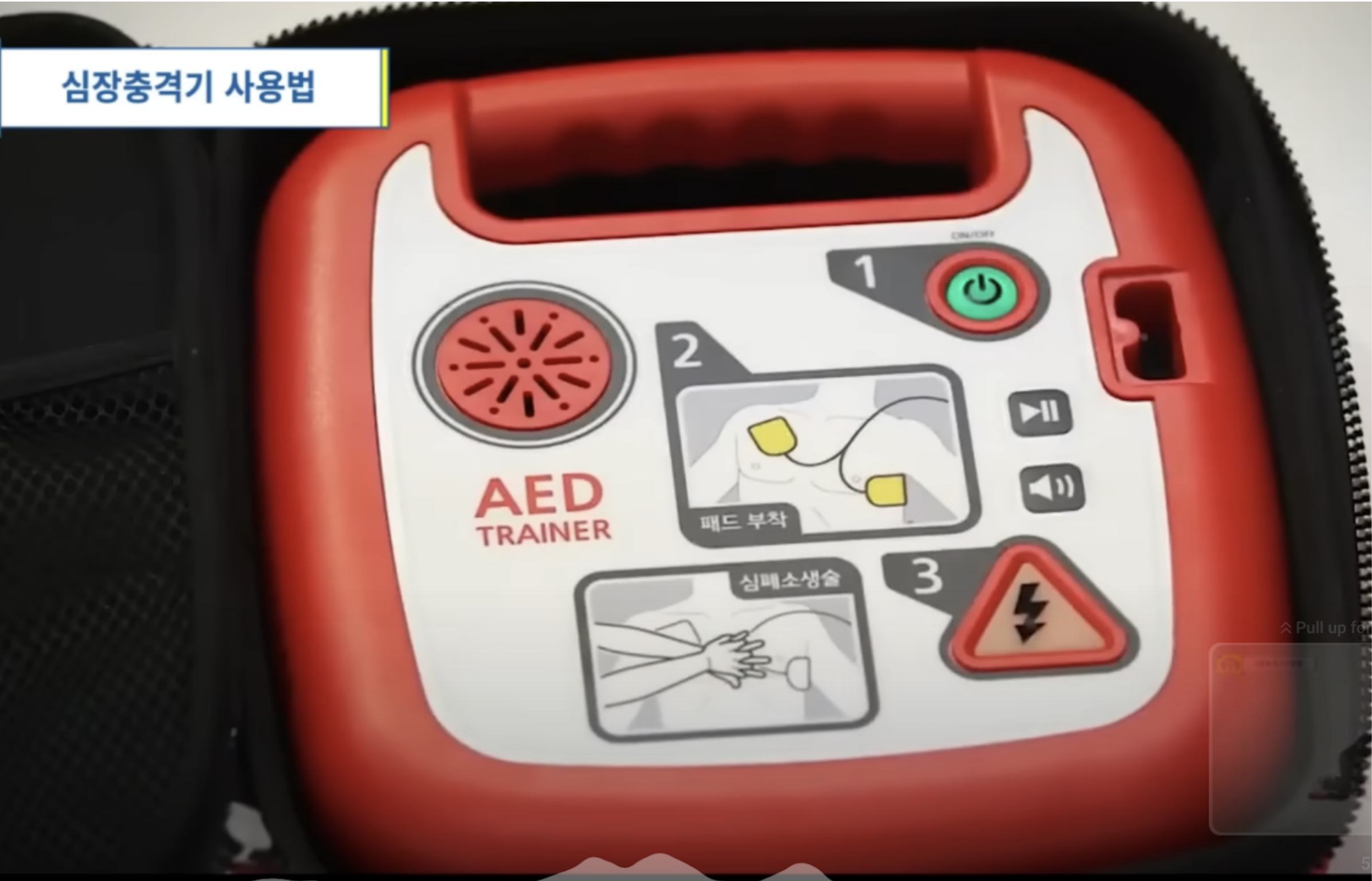}
\end{figure}

\textbf{Model Answer:} The green button next to number 1

\pointfour{\noindent\textit{Rationale: Tests visual search and identification of a specific button on a complex AED panel. Additionally, the AED in the video was shown at approximately -45 degrees rotation, testing whether participants could recognize the same device across different orientations.}}

\subsubsection*{Step 3}

\noindent\textbf{Q7.} Answer O/X (True/False). \label{appendix-Q7-2}
\begin{enumerate}
  \renewcommand{\labelenumi}{(\arabic{enumi})}
  \item AED pads have diagrams on them, so you can simply follow the pictures to place them. (\textbf{O}/X)
  \item It is better if someone continues chest compressions even while attaching the pads. (\textbf{O}/X)
\end{enumerate}

\pointfour{\noindent\textit{Rationale: (1) The video shows the diagrams on the pads visually without explicit narration. This tests whether participants could infer this information from visual cues alone. (2) This information is also not explicitly stated. Following Q5, this re-examines whether participants understood the importance of continuous chest compressions.}}

\vspace{0.5em}
\noindent\textbf{Q8.} In the video, the patient's clothes were removed before attaching the AED pads. Is it necessary to remove clothing to place the pads?
\begin{itemize}
  \item (A) As long as the clothes aren't too thick, it's fine.
  \item \textbf{(B) Yes. Clothes must be removed, and the pads must be attached directly to the skin.}
  \item (C) Attaching them over clothes works just as effectively.
  \item (D) It's okay as long as the clothes aren't wet.
\end{itemize}

\pointfour{\noindent\textit{Rationale: The video omits explicit instruction about removing clothing, only showing an undressed mannequin. Tests whether participants could infer this requirement from implicit visual presentation.}}

\vspace{0.5em}
\begin{figure}[H]
    \centering
    \includegraphics[width=0.8\linewidth]{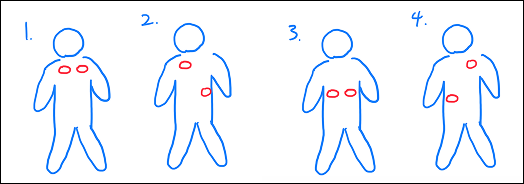}
\end{figure}
\noindent\textbf{Q9.} Which of the following illustrations shows the correct placement of the AED pads? \label{appendix-Q9}

\pointfour{\noindent\textit{Rationale: The video shows pad placement through both diagrams on the pads and a demonstration on the mannequin. Tests spatial understanding of pad positioning from video content, while also probing prior knowledge to assess spatial perception ability.}}

\subsubsection*{Step 4}

\noindent\textbf{Q10.} While the AED is analyzing the heart rhythm, who must move away from the patient?
\begin{itemize}
  \item (A) No one needs to move away.
  \item (B) Only the person performing CPR.
  \item (C) Only the person operating the AED.
  \item \textbf{(D) Everyone near the patient.}
\end{itemize}

\pointfour{\noindent\textit{Rationale: The video delivers this instruction through multiple channels: on-screen text (``No contact with patient during analysis and defibrillation''), visual demonstration of the CPR person stepping back, and narration explaining everyone must move away. Tests whether multi-channel information was properly integrated.}}

\vspace{0.5em}
\noindent\textbf{Q11.} Answer O/X (True/False). When people need to move away from the patient, the AED also gives a voice prompt instructing them to step away. (\textbf{O}/X) \label{appendix-Q11}

\pointfour{\noindent\textit{Rationale: The video implicitly conveys that the AED provides voice guidance through the phrase ``follow the voice instructions.'' Tests whether participants recognized this implicit information without prior knowledge.}}

\vspace{0.5em}
\noindent\textbf{Q12.} From the video, identify which voice prompts are coming from the AED.

\pointfour{\noindent\textit{Rationale: Tests whether participants could distinguish the AED's machine voice from the narrator (both female voices, potentially confusing). Participants could identify the AED voice through: (1) detecting the subtle mechanical quality of the voice, or (2) recognizing implicit caption cues (green heart symbol indicating AED speech).}}

\subsubsection*{Step 5}

\noindent\textbf{Q13.} Answer O/X (True/False). \label{appendix-Q13}
\begin{enumerate}
  \renewcommand{\labelenumi}{(\arabic{enumi})}
  \item If a shock is needed, the AED announces, ``Shock required.'' (\textbf{O}/X)
  \item If you don't move away quickly after the AED says ``Shock required. Please step away from the patient,'' you may also receive an electric shock. (\textbf{O}/X)
  \begin{figure}[H]
      \centering
      \includegraphics[width=0.8\linewidth]{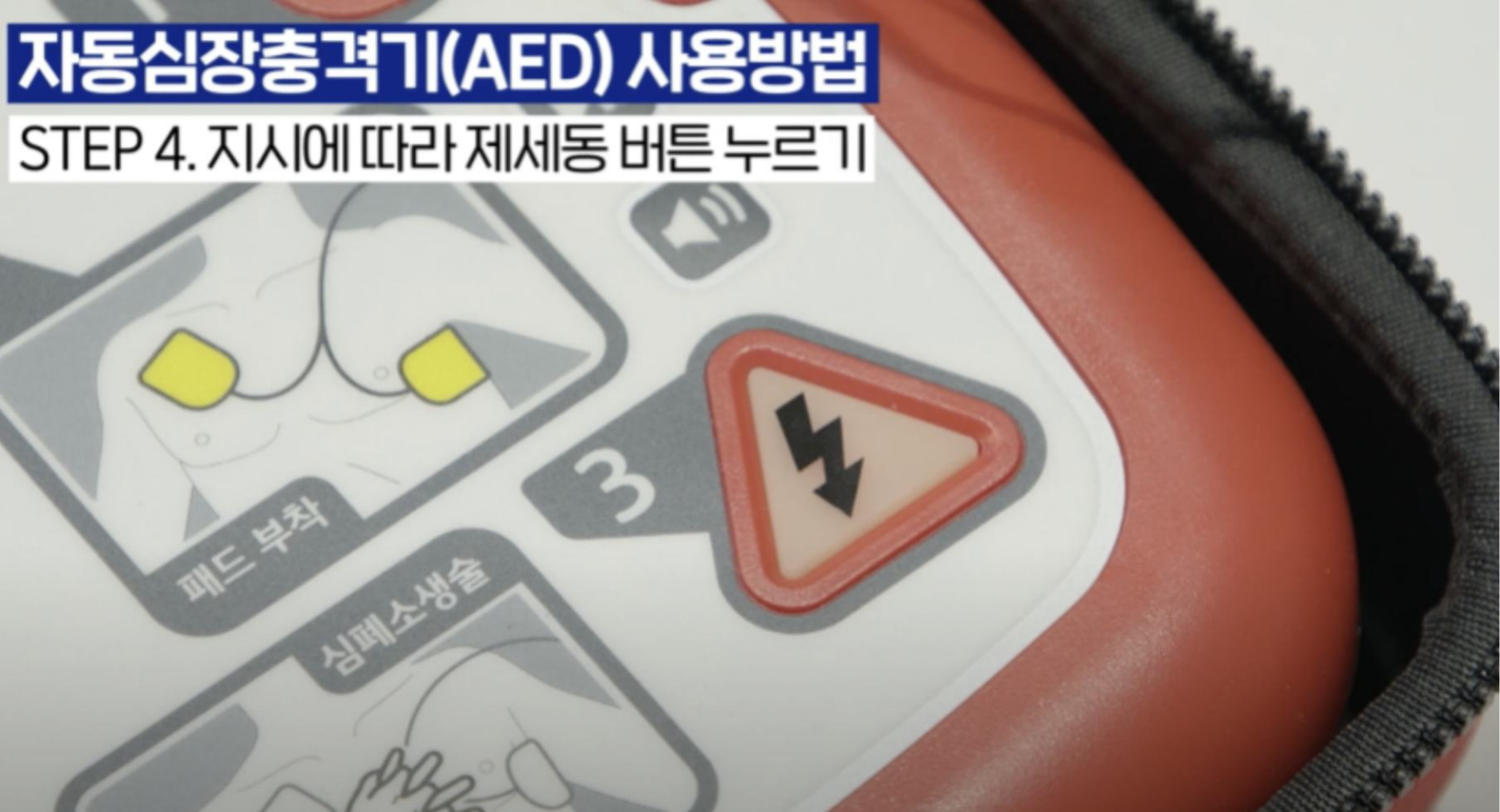}
  \end{figure}
  \item The lightning-shaped button in the picture is called the ``Defibrillation button.'' (O/\textbf{X})
  \item Pressing the lightning button immediately delivers an electric shock through the pads. (\textbf{O}/X)
  \item When the AED says, ``Shock required. Please step away from the patient,'' an automatic shock is delivered a few seconds later, so you must step away. (O/\textbf{X})
  \item The shock button blinks, and you should press it once it stops blinking and remains lit. (O/\textbf{X})
\end{enumerate}

\pointfour{\noindent\textit{Rationale: This multi-item question probes several aspects: (1) whether participants recognized the AED audio announcement; (2) understanding that shock does not occur automatically---the video transitions to the AED panel without explicit discourse markers (e.g., ``then'' or ``next'') and instructs to ``press the blinking button,'' which may cause confusion; (3) whether participants recognized the term ``defibrillation button''; (4) whether participants could infer that pressing the button delivers a shock, as the video does not visualize the electric shock; (5) a variation of item 2 to verify accurate understanding; (6) whether participants correctly understood the simultaneous audio instruction and visual demonstration of pressing the blinking button.}}

\vspace{0.5em}
\noindent\textbf{Q14.} What should you do immediately after pressing the lightning button?
\begin{itemize}
  \item (A) Check if the patient is breathing.
  \item (B) Wait until the AED gives the next instruction.
  \item \textbf{(C) Resume chest compressions (CPR).}
  \item (D) Turn off the AED.
\end{itemize}

\pointfour{\noindent\textit{Rationale: After showing the button press, the video provides narration instructing to ``immediately resume chest compressions after defibrillation.'' Tests whether participants understood this sequential instruction across scene transitions.}}

\subsubsection*{Outro}

\noindent\textbf{Q15.} Which of the following correctly summarizes the four steps of AED use in the right order?
\begin{itemize}
  \item (A) Turn on the power $\rightarrow$ Attach the pads $\rightarrow$ Call 119 $\rightarrow$ Deliver the shock (O)
  \item \textbf{(B) Turn on the power $\rightarrow$ Attach the pads $\rightarrow$ Step away from the patient $\rightarrow$ (If necessary) Deliver the shock}
  \item (C) Attach the pads $\rightarrow$ Deliver the shock $\rightarrow$ Perform CPR $\rightarrow$ Turn off the power
  \item (D) Perform CPR $\rightarrow$ Turn on the power $\rightarrow$ Attach the pads $\rightarrow$ Check the pulse
\end{itemize}

\pointfour{\noindent\textit{Rationale: Tests overall comprehension and retention of the complete procedural sequence presented in the video.}}

\end{document}